\documentclass[10pt,a4paper]{article}
\usepackage{amssymb}

\usepackage{graphicx}
\usepackage{amsmath}
\usepackage{makeidx}
\usepackage{indentfirst}

\newcounter{resultnum}[section]

\newcounter{conclusionnum}[section]

\newcounter{conditionnum}[section]

\newcounter{conjecturenum}[section]

\newcounter{examplenum}[section]

\newcounter{exercisenum}[section]

\newcounter{lemmanum}[section]

\newcounter{notationnum}[section]
\newtheorem{theorem}{Theorem}[section]

\newcounter{theoremnum}[section]
\newtheorem{definition}{Definition}[section]

\newcounter{definitionnum}[section]
\newtheorem{corollary}{Corollary}[section]

\newcounter{corollarynum}[section]
\newtheorem{remark}{Remark}[section]

\newcounter{remarknum}[section]
\newtheorem{proposition}{Proposition}[section]

\newcounter{propositionnum}[section]

\newcounter{acknowledgementnum}[section]

\newcounter{algorithmnum}[section]

\newcounter{axiomnum}[section]

\newcounter{casenum}[section]

\newcounter{claimnum}[section]

\newcounter{summarynum}[section]

\newcounter{problemnum}[section]
\newenvironment{proof}[1][]{\textbf{Proof.} }{}

\newcommand{ \C} {\mbox{\rm I$\!$C}}

\newcommand{ \I} {\mbox{\rm I$\!$I}}
\newcommand{ \D} {\mbox{\rm I$\!$\bf{D}}}

\begin{document}

\title{ Riemann--Finsler and Lagrange Gerbes\\ and
 the Atiyah--Singer Theorems}
 
 \author{\textbf{Sergiu I. Vacaru} \thanks{%
 sergiu.vacaru@uaic.ro, Sergiu.Vacaru@gmail.com} \\
\textsl{\small University "Al. I. Cuza" Ia\c si, Science Department,} \\
\textsl{\small 54 Lascar Catargi street, Ia\c si, Romania, 700107 }}
  
\date{January 10, 2012}
\maketitle

\begin{abstract}
In this paper, nonholonomic gerbes will be naturally derived for manifolds
and vector bundle spaces provided with nonintegrable distributions (in
brief, nonholonomic spaces). An important example of such gerbes is related
to distributions defining nonlinear connection (N--connection) structures.
They geometrically unify and develop the concepts of Riemann--Cartan
manifolds and Lagrange--Finsler spaces. The obstruction to the existence of
a spin structure on nonholonomic spaces is just the second Stiefel--Whitney
class, defined by the cocycle associated to a $\mathbb{Z} /2$ gerbe, which
is called the nonholonomic spin gerbe. The nonholonomic gerbes are
canonically endowed with N--connection, Sasaki type metric, canonical linear
connection connection and (for odd dimension spaces) almost complex
structures. The study of nonholonomic spin structures and gerbes have both
geometric and physical applications. Our aim is to prove the Atiyah--Singer
theorems for such nonholonomic spaces.



2000 AMS Subject Classification:\

 55R65, 53C05, 53C27, 57R20, 57R22, 53B20, 70H99, 81T13, 83C60

\end{abstract}



\section{ Introduction}

Connections and curving on gerbes (abelian and, more recently, nonabelian
ones) play an important role in modern differential geometry and
mathematical physics. Gerbes enabled with connection structure were
introduced as a natural higher-order generalization of abelian bundles with
connection provided a new possible framework to generalized gauge theories.
They appeared in algebraic geometry \cite{gir,attal}, and were subsequently
developed by Brylinski \cite{bry},  see a review of results in \cite{brm}.
Bundles and gerbes and their higher generalizations ($n$--gerbes) can be
understood both in two equivalent terms of local geometry \ (local functions
and forms) and of non-local geometry (holonomies and parallel transports) %
\cite{bar91,cp94,mp}.

The first applications of gerbes formalism were considered for higher
Yang--Mills fields and gravity \cite{bae} and for a special case of a
topological quantum field theory \cite{p03,ps}. The approaches were renewed
following Hitchin \cite{hi99} with further applications in physics, for
instance, in investigating anomalies \cite{cmm}, new geometrical structures
in string theory \cite{wit} and Chern--Simons theory \cite{gom}.

A motivation for noncommutative gerbes \cite{abjs}, related to deformation
quantization \cite{kon,burw}, follows from the noncommutative description of
D--branes in the presence of topologically non--trivial background fields.
In a more general context, the geometry of commutative and noncommutative
gerbes may be connected to the nonholonomic frame method in (non)
commutative gauge realizations and generalizations of the Einstein gravity %
\cite{vncgg}, nonholonomic deformations with noncommutative and/or algebroid
symmetry \cite{vesnc} and to the geometry of Lagrange--Fedosov
nonholonomic manifolds \cite{esv}. Here, we note that a manifold is
nonholonomic (equivalently, anholonomic) if it is provided with a
nonintegrable global distribution. In our works, we restrict the
constructions to a subclass of such nonholonomic manifolds, or bundle
spaces, when their nonholonomic distribution defines a nonlinear connection
(in brief, N--connection) structure. We use the term of N--anholonomic
manifold for such spaces. The geometry of N--connections came from the Finsler
and Lagrange geometry (see, for instance, details in Ref. \cite{ma}, and
Refs. \cite{ehr,dl} related to the Ehresmann connection and geometrization
of classical mechanics and field theory). Nevertheless, the N--connection
structures have to be introduced in general relativity, string theory and on
Riemann--Cartan and/or noncommutative spaces and various types of
non--Riemannian spaces if generic off--diagonal metrics, nonholonomic frames
and genalized connections are introduced into consideration, see discussion
and references in \cite{vncl}.

The concept of N--anholonomic spaces unifies a large class of nonholono\-mic
manifolds and bundle spaces, nonholonomic Einstein, non--Riemanni\-an and
Lagrange--Finsler geometries which are present in modern gravity and string
theory, geometric mechanics and classical field theory and geometric
quantization formalism. N--anholonomic spaces are naturally provided with
certain canonical N--connection and linear connection structures, Sasaki
type metric, almost complex and/or almost sympletic structures induced
correspondingly by the Lagrange, or Finsler, fundamental functions, and for
gravitational models by the generic off--diagonal metric terms. The
N--connection curvatures and Riemannian curvatures are very useful to study
the topology of such manifolds.

The study of bundles of spinors on N--anholonomic spaces provides a number
of geometric and physical results. For instance, it was possible
to give a definition of nonholonomic spinor structures for Finsler spaces %
\cite{vfs}, to get a spinor interpretation of Lagrange and Hamilton spaces
(and their higher order extensions), to define N--anholonomic Dirac operators
in connection to noncommutative extensions of Finsler--Lagrange geometry
and to construct a number of exact solutions with nonholonomic solitons and
spinor interactions \cite{vhs,vv,vncl}. But every compact manifold (being an
holonomic or anhlolonomic one) is not spin. The obstruction to the existence of spin
structure, in general, of any nonholonomic spin structure of a nonholonomic
space, is the second Stiefel--Whitney class. This class is also the
classifying cocycle associated to a $\mathbb{Z} /2$ gerbe, which with
respect to nonholonomic manifolds is a nonholonomic gerbe. This way one can
be constructed a number of new examples of gerbes (Finsler and/or Lagrange
ones, parametrizing some higher symmetries of generic off--diagonal
solutions in Einstein and string/ brane gravity, with additional
noncommutative and/or algebroid symmetries).

The aim of this paper is to study the main geometric properties of
nonholonomic gerbes. We shall generalize the Lichnerowicz theorem and prove
Atiyah--Singer type theorems for nonholonomic gerbes. For trivial holonomic
manifolds, our results will transform into certain similar ones from Refs. %
\cite{aris,mursin} but not completely if there are considered
'non-perturbative' and nonlinear configurations as exact solutions in
gravity models, see details in \cite{vgh}.

We note that the problem of formulating and proof of Atiyah--Singer type
theorems for nonholonomic manifolds is not a trivial one. For instance, in
Ref. \cite{ver}, it is advocated the point that it is not possible to define
the concept of curvature for general nonholonomic manifolds. Without
curvatures, one can not be formulated any types of Atiyah--Singer theorems.
In Refs. \cite{ver,leit2} one proposed such definitions for
supermanifolds when supersymmetric structure is treated as nonholonomic
distribution. There is a long term history on defining torsions and
curvatures for various classes of nonholonomic manifolds (see, for instance,
Refs. \cite{nhm3,nhm1}). More recently, one concluded that such definitions
can be given by using the concept of N--connection structure at least for
Lagrange--Finsler and Hamilton--Cartan spaces \cite{mironnh1,mironnh2}. \ We
note that the problem of definition of curvatures was discussed and solved
(also by using the N--connection formalism) in modern approaches to the
geometry of noncommutative Riemann--Finsler or Einstein spaces and
generalizations \cite{vncl,vesnc}, Fedosov N--anholonomic manifolds %
\cite{esv}, as well in the works on nonholonomic Clifford structures and
spinors \cite{vfs,vhs,vv} and generalized Finsler superspaces \cite{vncsup}.
The Lichnerowicz type formula and Atiyah--Singer theorems to be proved here
for N--anholonomic spaces have strong relations to the mentioned classes of
manifolds and supermanifolds.

The article is organized as follows. In section 2 we recall the main results
on nonholonomic manifolds provided with N--connection structure,
consider some examples of such N--anholonomic spaces (generalized
Lagrange/Finsler spaces and Riemann--Cartan manifolds provided
with N--connections) and define the concept of nonholonomic gerbe.
Section 3 is a study of nonhlonomic Clifford gerbes: we consider lifts
and nonholonomic vector gerbes and study pre--Hilbertian and scalar structures
and define distinguished (by the N--connection structure) linear connections on
N--anholonomic gerbes and construct the characteristic classes. Section 4 is
devoted to operators and symbols on nonholnomic gerbes. In section 5  we present
a $K$--theory framework for N--anholonomic manifolds and gerbes. We define
elliptic operators and index formulas adapted to the N--connection structure.
There are proven the Atiyah--Singer theorems for the canonical d--connection
and N--connection structures. The results  are applied for a topological study
of N--anholonomic spinors and related Dirac operators. Appendix A outlines
the basic results on distinguished connections, torsions and curvatures for
N--anholonomic manifolds. Appendix B is an introduction into the geometry of
nonholonomic spinor structures and N--adapted spin connections.

\vskip5pt
{\bf Acknowledgments:} This paper contains a series of results elaborated during  long term visits at CSIC, Madrid, Spain and Fields Institute, Toronto, Canada. Some important ideas and theorems were communicated in a review lecture at the VII-th International Conference  Finsler Extensions of Relativity Theory - FERT 2011,  29 Aug - 4 Sep 2011, Brasov - Romania,  September 2, 2011. Author is grateful to  M. Anastasiei, G. Munteanu and D. Pavlov,  N. Voicu and M. Neagu for kind support of his participation, hospitality  and important discussions. The research in this paper is partially supported by the Program IDEI, PN-II-ID-PCE-2011-3-0256.

\section{Nonholonomic Manifolds and Gerbes}

The aim of this section is to outline some results from the geometry of
nonholonomic manifolds provided with N--connection structure and to
elaborate the notion of nonholonomic gerbes.

\subsection{The geometry of N--anholonomic spaces}

We consider $(n+m)$--dimensional manifold of necessary smoothly class $%
\mathbf{V}$ with locally fibred structure. A particular case is that of a
vector bundle, when we shall write $\mathbf{V=E}$ (where $\mathbf{E}$ is
the total space of a vector bundle $\pi :$ $\mathbf{E}\rightarrow M$ with
the base space $M).$ We denote by $\pi ^{\top }:T\mathbf{V}\rightarrow TM$ the
differential of a map $\pi :V^{n+m}\rightarrow V^{n}$ defined by fiber
preserving morphisms of the tangent bundles $T\mathbf{V}$ and $TM.$ The
kernel of $\pi ^{\top }$ is just the vertical subspace $v\mathbf{V}$ with a
related inclusion mapping $i:v\mathbf{V}\rightarrow T\mathbf{V}.$

\begin{definition}
\label{dnc}A nonlinear connection (N--connection) $\mathbf{N}$ on a manifold
$\mathbf{V}$ is defined by the splitting on the left of an exact sequence%
\begin{equation*}
0\rightarrow v\mathbf{V}\overset{i}{\rightarrow }T\mathbf{V}\rightarrow T%
\mathbf{V}/v\mathbf{V}\rightarrow 0,
\end{equation*}%
i. e. by a morphism of submanifolds $\mathbf{N:\ \ }T\mathbf{V}\rightarrow v%
\mathbf{V}$ such that $\mathbf{N\circ i}$ is the unity in $v\mathbf{V}.$
\end{definition}

In an equivalent form, we can say that a N--connection is defined by a
splitting to subspaces with a Whitney sum of conventional horizontal (h)
subspace, $\left( h\mathbf{V}\right) ,$ and vertical (v) subspace, $\left( v%
\mathbf{V}\right) ,$
\begin{equation}
T\mathbf{V}=h\mathbf{V}\oplus v\mathbf{V}  \label{whitney}
\end{equation}%
where $h\mathbf{V}$ is isomorphic to $M.$ In general, a distribution (\ref%
{whitney}) in nonintegrabe, i.e. nonholonomic (equivalently, anholonomic).
In this case, we deal with nonholonomic manfiolds/ spaces.

\begin{definition}
A manifold \ $\mathbf{V}$ is called N--anholonomic if on the tangent space $T%
\mathbf{V}$ it is defined a local (nonintegrable) distribution (\ref{whitney}%
), i.e. $\mathbf{V}$ is N--anholonomic if it is enabled with a N--connection
structure.
\end{definition}

Locally, a N--connection is defined by its coefficients $N_{i}^{a}(u),$
\begin{equation*}
\mathbf{N}=N_{i}^{a}(u)dx^{i}\otimes \partial _{a}
\end{equation*}%
where the local coordinates (in general, abstract ones both for holonomic
and nonholonomic variables) are split in the form $u=(x,y),$ or $u^{\alpha
}=\left( x^{i},y^{a}\right) ,$ where $i,j,k,\ldots =1,2,\ldots ,n$ and $%
a,b,c,\ldots =n+1,n+2,\ldots ,n+m$ when $\partial _{i}=\partial /\partial
x^{i}$ and $\partial _{a}=\partial /\partial y^{a}.$ The well known class of
linear connections consists on a particular subclass with the coefficients
being linear on $y^{a},$ i.e., $N_{i}^{a}(u)=\Gamma _{bj}^{a}(x)y^{b}.$

A N--connection is characterized by its N--connection curvature (the
Nijenhuis tensor)
\begin{equation*}
\mathbf{\Omega }=\frac{1}{2}\Omega _{ij}^{a}dx^{i}\wedge dx^{j}\otimes
\partial _{a},
\end{equation*}%
with the N--connection curvature coefficients
\begin{equation}
\Omega _{ij}^{a}=\delta _{\lbrack j}N_{i]}^{a}=\delta _{j}N_{i}^{a}-\delta
_{i}N_{j}^{a}=\partial _{j}N_{i}^{a}-\partial
_{i}N_{j}^{a}+N_{i}^{b}\partial _{b}N_{j}^{a}-N_{j}^{b}\partial
_{b}N_{i}^{a}.  \label{ncurv}
\end{equation}

Any N--connection $\mathbf{N}=N_{i}^{a}(u)$ induces a N--adapted frame
(vielbein) structure
\begin{equation}
\mathbf{e}_{\nu }=\left( e_{i}=\partial _{i}-N_{i}^{a}(u)\partial
_{a},e_{a}=\partial _{a}\right) ,  \label{dder}
\end{equation}%
and the dual frame (coframe) structure%
\begin{equation}
\mathbf{e}^{\mu }=\left( e^{i}=dx^{i},e^{a}=dy^{a}+N_{i}^{a}(u)dx^{i}\right)
.  \label{ddif}
\end{equation}%
The vielbeins (\ref{ddif}) satisfy the nonholonomy (equivalently,
anholonomy) relations
\begin{equation}
\lbrack \mathbf{e}_{\alpha },\mathbf{e}_{\beta }]=\mathbf{e}_{\alpha }%
\mathbf{e}_{\beta }-\mathbf{e}_{\beta }\mathbf{e}_{\alpha }=W_{\alpha \beta
}^{\gamma }\mathbf{e}_{\gamma }  \label{anhrel}
\end{equation}%
with (antisymmetric) nontrivial anholonomy coefficients $W_{ia}^{b}=\partial
_{a}N_{i}^{b}$ and $W_{ji}^{a}=\Omega _{ij}^{a}.$\footnote{%
One preserves a relation to our previous denotations \cite{vfs,vhs} if
we consider that $\mathbf{e}_{\nu }=(e_{i},e_{a})$ and $\mathbf{e}^{\mu
}=(e^{i},e^{a})$ are, respectively, the former $\delta _{\nu }=\delta
/\partial u^{\nu }=(\delta _{i},\partial _{a})$ and $\delta ^{\mu }=\delta
u^{\mu }=(d^{i},\delta ^{a})$ when emphasize that operators (\ref{dder}) and
(\ref{ddif}) define, correspondingly, the ``N--elongated'' partial
derivatives and differentials which are convenient for calculations on
N--anholonomic manifolds.}

The geometric constructions can be adapted to the N--connection structure:

\begin{definition}
A distinguished connection (d--connection) $\mathbf{D}$ on a
N--anho\-lo\-no\-mic manifold $\mathbf{V}$ is a linear connection conserving
under parallelism the Whitney sum (\ref{whitney}).
\end{definition}

In this work we use boldfaced symbols for the spaces and geometric objects
provided/adapted to a N--connection structure. For instance, a vector field $%
\mathbf{X}\in T\mathbf{V}$ \ is expressed $\mathbf{X}=(X,\ ^{-}X),$ or $%
\mathbf{X}=X^{\alpha }\mathbf{e}_{\alpha }=X^{i}e_{i}+X^{a}e_{a},$ where $%
X=X^{i}e_{i}$ and $^{-}X=X^{a}e_{a}$ state, respectively, the irreducible
(adapted to the N--connection structure) horizontal (h) and vertical (v)
components of the vector (which following Refs. \cite{ma} is called a
distinguished vectors, in brief, d--vector). In a similar fashion, the
geometric objects on $\mathbf{V}$ like tensors, spinors, connections, ...
are called respectively d--tensors, d--spinors, d--connections if they are
adapted to the N--connection splitting.

One can introduce the d--connection 1--form%
\begin{equation*}
\mathbf{\Gamma }_{\ \beta }^{\alpha }=\mathbf{\Gamma }_{\ \beta \gamma
}^{\alpha }\mathbf{e}^{\gamma },
\end{equation*}%
when the N--adapted components of d-connection $\mathbf{D}_{\alpha }=(%
\mathbf{e}_{\alpha }\rfloor \mathbf{D})$ are computed following formulas
\begin{equation}
\mathbf{\Gamma }_{\ \alpha \beta }^{\gamma }\left( u\right) =\left( \mathbf{D%
}_{\alpha }\mathbf{e}_{\beta }\right) \rfloor \mathbf{e}^{\gamma },
\label{cond1}
\end{equation}%
where ''$\rfloor "$ denotes the interior product. This allows us to define in
standard form the torsion
\begin{equation}
\mathcal{T}^{\alpha }\doteqdot \mathbf{De}^{\alpha }=d\mathbf{e}^{\alpha
}+\Gamma _{\ \beta }^{\alpha }\wedge \mathbf{e}^{\beta }  \label{tors}
\end{equation}%
and curvature%
\begin{equation}
\mathcal{R}_{\ \beta }^{\alpha }\doteqdot \mathbf{D\Gamma }_{\beta }^{\alpha
}=d\mathbf{\Gamma }_{\beta }^{\alpha }-\Gamma _{\ \beta }^{\gamma }\wedge
\mathbf{\Gamma }_{\ \gamma }^{\alpha }.  \label{curv}
\end{equation}

There are certain preferred d--connection structures on N--anholonomic
manifolds (see local formulas in Appendix and Refs. \cite{ma,vesnc,esv}, for
details on computation the components of torsion and curvatures for various
classes of d--connections).

\subsection{Examples of N--anholonomic spaces:}

We show how the N--connection geometries can be naturally derived from
Lagrange--Finsler geometry and in gravity theories.

\subsubsection{Lagrange--Finsler geometry}

Such geometries are usually modelled on tangent bundles \cite{ma} but it is
possible to define such structures on general N--anholonomic manifolds, in
particular in (pseudo) Riemannian and Riemann--Cartan geometry if
nonholonomic frames are introduced into consideration \cite{vncl,vncgg}.
In the first approach the N--anholonomic manifold $\mathbf{V}$ is just a
tangent bundle $(TM,\!\pi ,\!M),$ where $M$ is a $n$--dimensional base
manifold, $\pi $ is a surjective projection and $TM$ is the total space. One
denotes by $\widetilde{TM}=TM\backslash \{0\}$ where $\{0\}$ means the null
section of map $\pi .$

A differentiable Lagrangian $L(x,y),$ i. e. a fundamental Lagrange function,
is defined by a map $L:(x,y)\in TM\rightarrow L(x,y)\in \mathbb{R}$ of class
$\mathcal{C}^{\infty }$ on $\widetilde{TM}$ and continuous on the null
section $0:M\rightarrow TM$ of $\pi .$ For simplicity, we consider any
regular Lagrangian with nondegenerated Hessian
\begin{equation}
\ ^{L}g_{ij}(x,y)=\frac{1}{2}\frac{\partial ^{2}L(x,y)}{\partial
y^{i}\partial y^{j}}  \label{lqf}
\end{equation}%
when $rank\left| g_{ij}\right| =n$ on $\widetilde{TM}$ and the left up ''L''
is an abstract label pointing that the values are defined by the Lagrangian $%
L.$

\begin{definition}
A Lagrange space is a pair $L^{n}=\left[ M,L(x,y)\right] $ with $\ \
^{L}g_{ij}(x,y)$ being of constant signature over $\widetilde{TM}.$
\end{definition}

The notion of Lagrange space was introduced by J. Kern \cite{kern} and
elaborated in details in Ref. \cite{ma} as a natural extension of Finsler
geometry.

By straightforward calculations, one can be proved the fundamental results:

\begin{enumerate}
\item The Euler--Lagrange equations%
\begin{equation*}
\frac{d}{d\tau }\left( \frac{\partial L}{\partial y^{i}}\right) -\frac{%
\partial L}{\partial x^{i}}=0
\end{equation*}%
where $y^{i}=\frac{dx^{i}}{d\tau }$ for $x^{i}(\tau )$ depending on
parameter $\tau ,$ are equivalent to the ``nonlinear'' geodesic equations
\begin{equation*}
\frac{d^{2}x^{i}}{d\tau ^{2}}+2G^{i}(x^{k},\frac{dx^{j}}{d\tau })=0
\end{equation*}%
defining paths of the canonical semispray%
\begin{equation*}
S=y^{i}\frac{\partial }{\partial x^{i}}-2G^{i}(x,y)\frac{\partial }{\partial
y^{i}}
\end{equation*}%
where
\begin{equation*}
2G^{i}(x,y)=\frac{1}{2}\ ^{L}g^{ij}\left( \frac{\partial ^{2}L}{\partial
y^{i}\partial x^{k}}y^{k}-\frac{\partial L}{\partial x^{i}}\right)
\end{equation*}%
with $^{L}g^{ij}$ being inverse to (\ref{lqf}).

\item There exists on $\widetilde{TM}$ a canonical N--connection $\ $%
\begin{equation}
\ ^{L}N_{j}^{i}=\frac{\partial G^{i}(x,y)}{\partial y^{i}},  \label{cncl}
\end{equation}%
defined by the fundamental Lagrange function $L(x,y),$ prescribing
nonholonomic frame structures of type (\ref{dder}) and (\ref{ddif}), $\ ^{L}%
\mathbf{e}_{\nu }=(e_{i},\ ^{-}e_{k})$ and $\ ^{L}\mathbf{e}^{\mu }=(e^{i},\
^{-}e^{k}).$ \footnote{%
On the tangent bundle the indices related to the base space run the same
values as those related to fibers: we can use the same symbols but have to
distinguish like $^{-}e_{k}$ certain irreducible v--components with respect
to, (or for) N--adapted bases and co--bases.}

\item The canonical N--connection (\ref{cncl}), defining $\ ^{-}e_{i},$
induces naturally an almost complex structure $\mathbf{F}:\chi (\widetilde{TM%
})\rightarrow \chi (\widetilde{TM}),$ where $\chi (\widetilde{TM})$ denotes
the module of vector fields on $\widetilde{TM},$%
\begin{equation*}
\mathbf{F}(e_{i})=\ ^{-}e_{i}\mbox{ and }\mathbf{F}(\ ^{-}e_{i})=-e_{i},
\end{equation*}%
when
\begin{equation}
\mathbf{F}=\ ^{-}e_{i}\otimes e^{i}-e_{i}\otimes \ ^{-}e^{i}  \label{acs1}
\end{equation}%
satisfies the condition $\mathbf{F\rfloor \ F=-I,}$ i. e. $F_{\ \ \beta
}^{\alpha }F_{\ \ \gamma }^{\beta }=-\delta _{\gamma }^{\alpha },$ where $%
\delta _{\gamma }^{\alpha }$ is the Kronecker symbol and ``$\mathbf{\rfloor }
$'' denotes the interior product.

\item On $\widetilde{TM},$ there is a canonical metric structure%
\begin{equation}
\ ^{L}\mathbf{g}=\ ^{L}g_{ij}(x,y)\ e^{i}\otimes e^{j}+\ ^{L}g_{ij}(x,y)\ \
^{-}e^{i}\otimes \ ^{-}e^{j}  \label{slm}
\end{equation}%
constructed as a Sasaki type lift from $M.$

\item There is also a canonical d--connection structure $\widehat{\mathbf{%
\Gamma }}_{\ \alpha \beta }^{\gamma }$ defined only by the components of $%
^{L}N_{j}^{i}$ and $^{L}g_{ij},$ i.e. by the \ coefficients of metric (\ref%
{slm}) which in its turn is induced by a regular Lagrangian. The
d--connection $\widehat{\mathbf{\Gamma }}_{\ \alpha \beta }^{\gamma }$ is
metric compatible and with vanishing $h$- and $v$--torsions.\ Such a
d--connection contains also nontrivial torsion components induced by the
nonholonomic frame structure, see Proposition \ref{pcdc} and formulas (\ref%
{candcon}) in Appendix. The canonical d--connection is the ''simplest''
N--adapted linear connection related by the ''non N--adapted'' Levi--Civita
connection by formulas (\ref{cdc}).
\end{enumerate}

We can conclude that any regular Lagrange mechanics can be geometrized as
an almost K\"{a}hler space with N--connection distribution, see \cite{ma,esv}.
 For the Lagrange--K\"{a}hler (nonholonomic) spaces, the fundamental
geometric structures (semispray, N--connection, almost complex structure and
canonical metric on $\widetilde{TM})$ are defined by the fundamental
Lagrange function $L(x,y).$

For applications in optics of nonhomogeneous media and gravity (see, for
instance, Refs. \cite{ma}) one considers metrics of type $g_{ij}\sim
e^{\lambda (x,y)}\ ^{L}g_{ij}(x,y)\ $\ which can not be derived from a
mechanical Lagrangian but from an effective ''energy'' function. In the
so--called generalized Lagrange geometry one considers Sasaki type metrics (%
\ref{slm}) with any general coefficients both for the metric and
N--connection.

\begin{remark}
A Finsler space is defined by a fundamental Finsler function $F(x,y),$ being
homogeneous of type $F(x,\lambda y)=\lambda F(x,y),$ for nonzero $\lambda
\in \mathbb{R},$ may be considered as a particular case of Lagrange geometry
when $L=F^{2}.$
\end{remark}

Now we show how N--anholonomic configurations can defined in gravity
theories. In this case, it is convenient to work on a general manifold $%
\mathbf{V},\dim \mathbf{V}=n+m$ with global splitting, instead of the
tangent bundle $\widetilde{TM}.$

\subsubsection{N--connections and gravity}

Let us consider a metric structure on $\mathbf{V}$ with the coefficients
defined with respect to a local coordinate basis $du^{\alpha }=\left(
dx^{i},dy^{a}\right) ,$%
\begin{equation*}
\mathbf{g}=\underline{g}_{\alpha \beta }(u)du^{\alpha }\otimes du^{\beta }
\end{equation*}%
with
\begin{equation}
\underline{g}_{\alpha \beta }=\left[
\begin{array}{cc}
g_{ij}+N_{i}^{a}N_{j}^{b}h_{ab} & N_{j}^{e}h_{ae} \\
N_{i}^{e}h_{be} & h_{ab}%
\end{array}%
\right] .  \label{ansatz}
\end{equation}%
In general, such a metric (\ref{ansatz})\ is generic off--diagonal, i.e it
can not be diagonalized by any coordinate transforms. We not that $%
N_{i}^{a}(u)$ in our approach are any general functions. They my be
identified with some gauge potentials in Kaluza--Klein models if the
corresponding symmetries and compactifications of coordinates $y^{a}$ are
considered, see review \cite{over}. Performing a frame transform
\begin{equation*}
\mathbf{e}_{\alpha }=\mathbf{e}_{\alpha }^{\ \underline{\alpha }}\partial _{%
\underline{\alpha }}\mbox{ and }\mathbf{e}_{\ }^{\beta }=\mathbf{e}_{\
\underline{\beta }}^{\beta }du^{\underline{\beta }}.
\end{equation*}%
with coefficients

\begin{eqnarray}
\mathbf{e}_{\alpha }^{\ \underline{\alpha }}(u) &=&\left[
\begin{array}{cc}
e_{i}^{\ \underline{i}}(u) & N_{i}^{b}(u)e_{b}^{\ \underline{a}}(u) \\
0 & e_{a}^{\ \underline{a}}(u)%
\end{array}%
\right] ,  \label{vt1} \\
\mathbf{e}_{\ \underline{\beta }}^{\beta }(u) &=&\left[
\begin{array}{cc}
e_{\ \underline{i}}^{i\ }(u) & -N_{k}^{b}(u)e_{\ \underline{i}}^{k\ }(u) \\
0 & e_{\ \underline{a}}^{a\ }(u)%
\end{array}%
\right] ,  \label{vt2}
\end{eqnarray}%
we write equivalently the metric in the form
\begin{equation}
\mathbf{g}=\mathbf{g}_{\alpha \beta }\left( u\right) \mathbf{e}^{\alpha
}\otimes \mathbf{e}^{\beta }=g_{ij}\left( u\right) e^{i}\otimes
e^{j}+h_{ab}\left( u\right) \ ^{-}e^{a}\otimes \ ^{-}e^{b},  \label{metr}
\end{equation}%
where $g_{ij}\doteqdot \mathbf{g}\left( e_{i},e_{j}\right) $ and $%
h_{ab}\doteqdot \mathbf{g}\left( e_{a},e_{b}\right) $ \ and the vielbeins $%
\mathbf{e}_{\alpha }$ and $\mathbf{e}^{\alpha }$ are respectively of type (%
\ref{dder}) and (\ref{ddif}). We can consider a special class of manifolds
provided with a global splitting into conventional ``horizontal'' and
``vertical'' subspaces (\ref{whitney}) induced by the ``off--diagonal''
terms $N_{i}^{b}(u)$ and prescribed type of nonholonomic frame structure.

If the manifold $\mathbf{V}$ is (pseudo) Riemannian, there is a unique
linear connection (the Levi--Civita connection) $\nabla ,$ which is metric, $%
\nabla \mathbf{g=0,}$ and torsionless, $\ ^{\nabla }T=0.$ Nevertheless, the
connection $\nabla $ is not adapted to the nonintegrable distribution
induced by $N_{i}^{b}(u).$ In this case, it is more convenient to work with
more general classes of linear connections (for instance, with the canonical
d--connection (\ref{candcon})) which are N--adapted but contain nontrivial
torsion coefficients because of nontrivial nonholonomy coefficients $%
W_{\alpha \beta }^{\gamma }$ (\ref{anhrel}).

For a splitting of a (pseudo) Riemannian--Cartan space of dimension $(n+m)$
( we considered also certain (pseudo) Riemannian configurations), the
Lagrange and Finsler type geometries were modelled by N--anholonomic
structures as exact solutions of gravitational field equations \cite%
{vncl,vncgg}.

\subsection{The notion of nonholonomic gerbes}

Let denote by $\mathbf{S}$ a sheaf of categories on a N--anholonomic
manifold $\mathbf{V},$ defined by a map of $\mathbf{U}\rightarrow \mathbf{S}(%
\mathbf{U}),$ where $\mathbf{U}$ is a open subset of $\mathbf{V,}$ with $%
\mathbf{S}(\mathbf{U}).$

\begin{definition}
A sheaf of categories $\mathbf{S}$ is called a nonholonomic gerbe if there
are satisfied the conditions:

\begin{enumerate}
\item There exists a map $r_{\mathbf{U}_{\widehat{1}}\mathbf{U}_{\widehat{2}%
}}:\mathbf{S}(\mathbf{U}_{\widehat{1}})\rightarrow \mathbf{S}(\mathbf{U}_{%
\widehat{2}})$ such that for superpositions of two such maps $r_{\mathbf{U}%
_{\widehat{1}}\mathbf{U}_{\widehat{2}}}\circ r_{\mathbf{U}_{\widehat{2}}%
\mathbf{U}_{\widehat{3}}}=r_{\mathbf{U}_{\widehat{1}}\mathbf{U}_{\widehat{3}%
}}$ for any inclusion $\mathbf{U}_{\widehat{1}}\rightarrow \mathbf{U}_{%
\widehat{2}}.$

\item It is satisfied the gluing condition for objects, i.e. for a covering
family $\cup _{\widehat{i}}\mathbf{U}_{\widehat{i}}$ of $\mathbf{U}$ and
objects $\mathbf{u}_{\widehat{i}}$ of $\mathbf{S}(\mathbf{U}_{\widehat{i}})$
for each $\widehat{i},$ when there are maps of type
\begin{equation*}
q_{\widehat{i}\widehat{j}}:r_{\mathbf{U}_{\widehat{i}}\cap \mathbf{U}_{%
\widehat{j}},\mathbf{U}_{\widehat{j}}}\left( \mathbf{u}_{\widehat{j}}\right)
\rightarrow r_{\mathbf{U}_{\widehat{i}}\cap \mathbf{U}_{\widehat{j}},\mathbf{%
U}_{\widehat{i}}}\left( \mathbf{u}_{\widehat{i}}\right)
\end{equation*}%
such that $q_{\widehat{i}\widehat{j}}q_{\widehat{j}\widehat{k}}=q_{\widehat{i%
}\widehat{k}},$ then there exists and object $\mathbf{u}\in \mathbf{S(U)}$
such that $r_{\mathbf{U}_{\widehat{i}},\mathbf{U}}\left( \mathbf{u}\right)
\rightarrow \mathbf{u}_{\widehat{i}}.$

\item It is satisfied the gluing condition for arrows, i.e. for any two
objects $\mathbf{P,Q}\in \mathbf{S}(\mathbf{V})$ the map
\begin{equation*}
\mathbf{U\rightarrow }Hom(r_{\mathbf{UV}}(\mathbf{P}),r_{\mathbf{UV}}(%
\mathbf{Q}))
\end{equation*}%
is a sheaf.\
\end{enumerate}
\end{definition}

This Definition is adapted to the N--connection structure (\ref{whitney})
and define similar objects and maps for h-- and v--subspaces of a
N--anholonomic manifold $\mathbf{V.}$ That why we use ''boldfaced'' symbols.

For certain applications is convenient to work with another sheaf $\mathbf{A}$
called the N--bund of the nonholonomic gerbe $\mathbf{S.}$ It is constructed
to satisfy the conditions:

\begin{itemize}
\item There is a covering N--adapted family $\left( \mathbf{U}_{\widehat{i}%
}\right) _{\widehat{i}\in I}$ of $\mathbf{V}$ such that the category of $%
\mathbf{S}(\mathbf{U}_{\widehat{i}})$ is not empty for each $\widehat{i}.$

\item For any $\mathbf{u}_{(1)},\mathbf{u}_{(2)}\in \mathbf{U\subset V,}$
there is a covering family $\left( \mathbf{U}_{\widehat{i}}\right) _{%
\widehat{i}\in I}$ of $\mathbf{U}$ such that $r_{\mathbf{U}_{\widehat{i}}%
\mathbf{U}}(\mathbf{u}_{(1)})$ and $r_{\mathbf{U}_{\widehat{i}}\mathbf{U}}(%
\mathbf{u}_{(2)})$ are isomorphic.

\item The N--bund is introduced as a family of isomorphisms $\mathbf{A(%
\mathbf{V})\doteqdot }Hom(\mathbf{u,}$ $\mathbf{u}),$ for each object $%
\mathbf{u}\in \mathbf{S}(\mathbf{U}),$ defined by a sheaf $\mathbf{A}$ in
groups, for which every arrow of $\mathbf{S}(\mathbf{U})$ is invertible and
such isomorphisms commute with the restriction maps.
\end{itemize}

For given families $\left( \mathbf{U}_{\widehat{i}}\right) _{\widehat{i}\in
I}$ of $\mathbf{V}$ and objects $\mathbf{u}_{\widehat{i}}$ of $\mathbf{S}(%
\mathbf{U}_{\widehat{i}}),$ we denote by $\mathbf{u}_{\widehat{i}_{1}...%
\widehat{i}_{k}}^{\widehat{i}}$ the element $r_{(\mathbf{U}_{\widehat{i}%
_{1}}\cap ...\cap \mathbf{U}_{\widehat{i}_{k}},\mathbf{U}_{\widehat{j}%
})}\left( \mathbf{u}_{\widehat{i}}\right) $ and by $\mathbf{U}_{\widehat{i}%
_{1}...\widehat{i}_{k}}$ the elements of the intersection $\mathbf{U}_{%
\widehat{i}_{1}}\cap ...\cap \mathbf{U}_{\widehat{i}_{k}}.$ The
N--connection structure distinguishes (d) $\mathbf{V}$ into h-- and
v--components, i.e. defines a local fiber structure when the geometric
objects transform into d--objects, for instance, d--vectors,
d--tensors,..... There are two possibilities for further constructions: a)
to consider the category of vector bundles over an open set $\mathbf{U}$ of
N--anholonomic manifold $\mathbf{V,}$ being the base space or b) to consider
such N--anholonomic vector bundles modelled as $\mathbf{V=E}$ with a base $%
M, $ where $\dim M=n$ and $\dim \mathbf{E=}n+m.$

\begin{definition}
\label{dnvg}{\quad } \newline
a) A N--anholonomic vector gerbe $\mathbf{C}_{NQ}$ is defined by the
category of vector bundles $\mathbf{S}(\mathbf{U})$ over $\mathbf{U\subset V}
$ with typical fiber the vector space $Q.$\newline
b) A nonholonomic gerbe $\mathbf{C}_{Nd}$ is a d--vectorial gerbe $\mathbf{S}%
(U)$ if and only if for the each open $U\subset M$ on the h--subspace $M$ of
$\mathbf{V}$ the set $\mathbf{S}(U)$ is a category of N--anholonomic
manifolds with h--base $M.$\newline
In both cases of nonholonomic gerbes a) and b) the maps between d--objects
are isomorphisms of N--anholonomic bundles/ manifolds adapted to the
N--connec\-ti\-on structures.
\end{definition}

Let us consider more precisely the case a) (the constructions for the case
b) being similar by substituting $\mathbf{U\rightarrow }U$\textbf{\ }and $%
\mathbf{V\rightarrow }M\mathbf{).}$ There is a covering family $\left(
\mathbf{U}_{\widehat{\alpha }}\right) _{\widehat{\alpha }\in I}$ of $\mathbf{%
V}$ and a commutative subgroup $H$ of the set of linear transforms $Gl(Q),$
such that there exit maps
\begin{eqnarray*}
q_{\widehat{\alpha }\widehat{\beta }}^{\prime } &:&\ \mathbf{U}_{\widehat{%
\alpha }}\cap \mathbf{U}_{\widehat{\beta }}\times Q\rightarrow \mathbf{U}_{%
\widehat{\alpha }}\cap \mathbf{U}_{\widehat{\beta }}\times Q, \\
q_{\widehat{\alpha }\widehat{\beta }}^{\prime } &:&\ (\mathbf{u}_{(1)},%
\mathbf{u}_{(2)})\rightarrow \left( \mathbf{u}_{(1)},q_{\widehat{\alpha }%
\widehat{\beta }}^{\prime }(\mathbf{u}_{(1)})\mathbf{u}_{(2)}\right)
\end{eqnarray*}%
when $c_{\widehat{\alpha }\widehat{\beta }\widehat{\gamma }}=q_{\widehat{%
\alpha }\widehat{\beta }}^{\prime }q_{\widehat{\beta }\widehat{\gamma }%
}^{\prime }q_{\widehat{\gamma }\widehat{\alpha }}^{\prime }$ is an $H$
2--Cech cocycle. Locally, such maps are parametrized by non--explicit
functions because of nonholonomic character of manifolds and subspaces under
consideration.

\section{Nonholonomic Clifford Gerbes}

\label{snclg}Let $\mathbf{V}$ be an N--anholnomic manifold of dimension $%
\dim \mathbf{V}=n+m.$ We denote by $O(\mathbf{V}),$ see applications and
references in \cite{vncgg}, the reduction of linear N--adapted frames which
defines the d--metric structure (\ref{metr}) of the $\mathbf{V.}$ The
typical fiber of $O(\mathbf{V})$ is $O(n+m)$ which with respect to
N--adapted frames splits into $O(n)\oplus O(m).$ There is  the exact
sequence%
\begin{equation*}
1\rightarrow \mathbb{Z}/2\rightarrow Spin(n+m)\rightarrow O(n+m)\rightarrow 1
\end{equation*}%
with two N--distinguished, respectively, h-- and v--components
\begin{eqnarray*}
1 &\rightarrow &\mathbb{Z}/2\rightarrow Spin(n)\rightarrow O(n)\rightarrow 1,
\\
1 &\rightarrow &\mathbb{Z}/2\rightarrow Spin(m)\rightarrow O(m)\rightarrow 1
\end{eqnarray*}%
where $Spin(n+m)$ is the universal covering of $O(n+m)$ splitting into $%
Spin(n)\oplus Spin(m)$ distinguished as the universal covering of $%
O(n)\oplus O(m).$ To such sequences, one can be associated a nonholnomic
gerbe with band $\mathbb{Z}/2$ and such that for each open set $\mathbf{%
U\subset V}$ it defined $Spin_{N}(\mathbf{U})$ as the category of $Spin$ $N$%
--anholonomic bundles over $\mathbf{U,}$ such spaces were studied in details
in Refs. \cite{vncl,vfs,vhs,vv}, see also Appendix \ref{sanssc}. The
classified cocycle of this N--anholonomic gerbe is defined by the second
Stiefel--Whitney class.

In a more general context, the N--anholonomic gerbe and $Spin_{N}(\mathbf{U}%
) $ are associated to a vectorial N--ahnolonomic gerbe called the Clifford
N--gerbe (in a similar form we can consider associated Clifford d--gerbe,
for the case b) of Definition \ref{dnvg}). This way one defines the category
$Cl_{N}(\mathbf{V})$ which for any open set $\mathbf{U\subset V,}$ one have
the category of objects being Clifford bundles provided with N--connection
structure associated to the objects of $Spin_{N}(\mathbf{U}).$ We can
consider such gerbes in terms of transition functions. Let $q_{\widehat{%
\alpha }\widehat{\beta }}^{\prime }$ be the transitions functions of the
bundle $O(\mathbf{V}).$ The N--connection distinguish them to couples of h-
and v--transition functions, i.e. $q_{\widehat{\alpha }\widehat{\beta }%
}^{\prime }=(q_{\widehat{i}\widehat{j}}^{\prime },q_{\widehat{a}\widehat{b}%
}^{\prime }).$ For such d--functions one can be considered elements $q_{%
\widehat{\alpha }\widehat{\beta }}=(q_{\widehat{i}\widehat{j}},q_{\widehat{a}%
\widehat{b}})$ acting correspondingly in $Spin(n+m)=\left(
Spin(n),Spin(m)\right) .$ Such elements act, by left multiplication,
correspondingly on $Cl(\mathbb{R}^{n+m})$ distinguished into $\left( Cl(%
\mathbb{R}^{n}),Cl(\mathbb{R}^{m})\right) .$ We denote by $s_{\widehat{%
\alpha }\widehat{\beta }}(\mathbf{u})=(s_{\widehat{i}\widehat{j}}(\mathbf{u}%
),s_{\widehat{a}\widehat{b}}(\mathbf{u}))$ the resulting automorphisms on
Clifford spaces. We conclude that the Clifford N--gerbe is defined by maps
\begin{equation*}
s_{\widehat{\alpha }\widehat{\beta }}:\mathbf{U}_{\widehat{\alpha }}\cap
\mathbf{U}_{\widehat{\beta }}\rightarrow Spin(n+m)
\end{equation*}%
distinguished with respect to N--adapted frames by couples
\begin{equation*}
s_{\widehat{i}\widehat{j}}:\mathbf{U}_{\widehat{i}}\cap \mathbf{U}_{\widehat{%
j}}\rightarrow Spin(n)\mbox{\ and \ }s_{\widehat{a}\widehat{b}}:\mathbf{U}_{%
\widehat{a}}\cap \mathbf{U}_{\widehat{b}}\rightarrow Spin(m).
\end{equation*}%
For trivial N--connections, such Clifford N--gerbes transform into the usual
Clifford gerbes defined in Ref. \cite{aris}. \footnote{%
We apply the ideas and results developed in that paper in order to
investigate N--anholonomic manifolds and gerbes.}

\subsection{Gerbes and lifts associated to d--vector bundles}

There are two classes of nonholonomic gerbes defined by lifting problems,
respectively, associated to a vector bundle $\mathbf{E}$ on a N--anholonomic
manifold $\mathbf{V}$ and/or associated just to $\mathbf{V}$ considering
that locally a such space posses a fibered structure distinguished by the
N--connection, see corresponding cases a) and b) in Definition \ref{dnvg}.

\subsubsection{Lifts and N--anholonomic vector gerbes}

Let us denote by $Q$ the typical fiber of a vector bundle $\mathbf{E}$ on $%
\mathbf{V}$ with associated principal bundle $Gl(Q).$ We suppose that this
bundle has a reduction $\mathbf{E}_{K}$ for a subgroup $K\subset Gl(Q)$ and
consider a central extension for a group $G$ when%
\begin{equation}
1\rightarrow H\rightarrow G\rightarrow K\rightarrow 1.  \label{ext1}
\end{equation}%
Such an extension defines a N--anholonomic gerbe $\mathbf{C}_{H}$ on $%
\mathbf{V}$ when for each open set $\mathbf{U}\subset \mathbf{V}$ the
objects of $\mathbf{C}_{H}(\mathbf{U})$ are $G$--principal bundles over $%
\mathbf{U}$ when the quotient by $H$ is the restriction of $\mathbf{E}_{K}$
to $\mathbf{U}.$

We consider the projection $\pi :G\rightarrow K$ and suppose that for
(existing) a representation $r:G\rightarrow Gl(Q^{\prime })$ and surjection $%
f:Q^{\prime }\rightarrow Q$ one can be defined a commutative Diagram 1,

\begin{figure}[tbph]
\begin{center}
\begin{picture}(140,80)\setlength{\unitlength}{1pt}%
\thinlines \put(12,60){$Q^{\prime }$}%
\put(100,60){$Q^{\prime }$}%
\put(12,0){$Q$}%
\put(100,0){${Q}$}%
\put(24,3){ \vector(1,0){65}}%
\put(16,54){\vector(0,-1){40}} %
\put(105,54) {\vector(0,-1){40}}%
\put(8,35){$f$}%
\put(50,12){$\underline{\pi (s)}$}%
\put(50,71){$\underline{r(s)}$}%
\put(108,35){$f$}%
\put(26,62){\vector(1,0){65}}%
\end{picture}
\end{center}
\caption{{Diagram 1}}
\end{figure}
%
%
For such cases it is defined a N--anholonomic vectorial gerbe $\mathbf{C}%
_{H,Q^{\prime }}$ on $\mathbf{V}$ when an object of $\mathbf{C}(\mathbf{U})$
is parametrized $\mathbf{e}_{U}\propto r$ where $\mathbf{e}_{U}$ is an
object of $\mathbf{C}_{H}(\mathbf{U}).$ Such constructions are adapted to
the N--connection structure on $\mathbf{U.}$ If $\left( \mathbf{U}_{%
\widetilde{\alpha }}\right) _{\widetilde{\alpha }\in I}$ is a trivialization
of $\mathbf{E,}$ with transition functions $q_{\widetilde{\alpha }\widetilde{%
\beta }}^{\prime }=(q_{\widetilde{i}\widetilde{j}}^{\prime },q_{\widetilde{a}%
\widetilde{b}}^{\prime }),$ we can define the maps $q_{\widetilde{\alpha }%
\widetilde{\beta }}:\mathbf{U}_{\widetilde{\alpha }}\cap \mathbf{U}_{%
\widetilde{\beta }}\rightarrow G$ over $q_{\widetilde{\alpha }\widetilde{%
\beta }}^{\prime }.$ This states that $\mathbf{C}_{H,Q^{\prime }}$ is
defined by $r(q_{\widetilde{\alpha }\widetilde{\beta }}).$

There is a natural scalar product defined on such N--anholonomic gerbes.
Its existence follows from the construction of Clifford N--gerbe $Cl(\mathbf{%
V})$ because the group $Spin$ is compact and its action on $Cl(\mathbb{R}%
^{n+m})$ preserves a scalar product which is distinguished by the
N--connection structure \cite{vncl,vfs,vhs,vv}. We can consider this scalar
product on each fiber of an object $\mathbf{e}_{U}$ of $Cl(\mathbf{U})$ and
define a Riemannian d--metric
\begin{equation*}
<,>_{e_{U}}=<,>_{he_{U}}+<,>_{ve_{U}}
\end{equation*}%
distinguished by the N--splitting into h- and v--components. The family of
such Riemannian d--metrics defines the Riemannian d--metric on the
N--anholonomic gerbe $Cl(\mathbf{V}).$ There is a canonical such structure
defined by the N--connection when $(g_{ij},h_{ab})$ in (\ref{metr}) are
taken to be some Euclidean ones but $N_{i}^{a}(\mathbf{u})$ are the
coefficients for a nontrivial N--connection.

It should be emphasized that the band has to be contained in a compact group
in order to preserve the Riemannian d--metric. In result, we can give the

\begin{definition}
A Riemannian d--metric on a N--anholonomic vector gerbe $\mathbf{C}_{NQ}$ is
given by a distinguished scalar product $%
<,>_{e_{U}}=(<,>_{he_{U}},<,>_{ve_{U}})$ on the vector bundle $\mathbf{e}%
_{U},$ defined for every object of $\mathbf{C}_{NQ}$ and preserved by
morphisms of such objects.
\end{definition}

We can define a global section of a N--anholonomic vector gerbe associated
to a 1--Cech N--adapted chain $q_{\widehat{\alpha }\widehat{\beta }}=(q_{%
\widehat{i}\widehat{j}},q_{\widehat{a}\widehat{b}})$ by considering a
covering space $\left( \mathbf{U}_{\widehat{\alpha }}\right) _{\widehat{%
\alpha }\in I}$ of $\mathbf{V}$ when for each element $\widehat{\alpha }$ of
$I,$ an object $\mathbf{e}_{\widehat{\alpha }}\in \mathbf{C}(\mathbf{U}_{%
\widehat{\alpha }}),$ a section $z_{\widehat{\alpha }}$ of $\ \mathbf{e}_{%
\widehat{\alpha }}$ and a family of morphisms $q_{\widehat{\alpha }\widehat{%
\beta }}:\ \mathbf{e}_{\widehat{\beta }}^{\widehat{\alpha }}\rightarrow \
\mathbf{e}_{\widehat{\alpha }}^{\widehat{\beta }}$ one has $z_{\widehat{%
\alpha }}=q_{\widehat{\alpha }\widehat{\beta }}(z_{\widehat{\beta }}).$ The
family of global sections $\mathbf{Z}\left( q_{\widehat{\alpha }\widehat{%
\beta }}\right) $ associated to $\left( q_{\widehat{\alpha }\widehat{\beta }%
}\right) _{\widehat{\alpha },\widehat{\beta }\in I}$ defining a d--vector
space. If $\mathbf{V}$ is compact and $I$ is finite, one can prove that $%
\mathbf{Z}\left( q_{\widehat{\alpha }\widehat{\beta }}\right) $ is not
empty. For such conditions, we can generalize for N--anholonomic vector
gerbes the Proposition 5 from Ref. \cite{aris},

\begin{proposition}
\label{pnvg}Let the N--anholonomic vector gerbe $\mathbf{C}_{NQ}$ is a
nonholonomic gerbe associated to the lifting problem defined by the
extension (\ref{ext1}) and the vector bundle $E$ and for a reprezentation $%
r:G\rightarrow Gl(Q^{\prime })$ the conditions of the Diagram 1 are
satisfied. Then for each $G$--chain $q_{\widehat{\alpha }\widehat{\beta }}$
and each element $\left( z_{\widehat{\alpha }}\right) _{\widehat{\alpha }\in
I}$ of the d--vector space of global sections $\mathbf{Z}\left( q_{\widehat{%
\alpha }\widehat{\beta }}\right) $ it is satisfied the condition that there
is a section $\mathbf{z}$ of $\mathbf{E}$ such that $\mathbf{z}_{|U_{%
\widehat{\alpha }}}=f\circ $ $\mathbf{z}_{\widehat{\alpha }}.$
\end{proposition}

\begin{proof}
It is similar to that for the usual vector bundles given in \cite{aris} but
it should be considered for both h-- and v--subspaces of $\mathbf{V}$ and $%
\mathbf{E}.$ In ''non--distinguished'' form, we can consider a global
section $\left( z_{\widehat{\alpha }}\right) _{\widehat{\alpha }\in I}$
associate to $\mathbf{Z}\left( q_{\widehat{\alpha }\widehat{\beta }}\right)
. $ One has $z_{\widehat{\alpha }}=q_{\widehat{\alpha }\widehat{\beta }}(z_{%
\widehat{\beta }})$ implying that $f(z_{\widehat{\alpha }})=f(z_{\widehat{%
\beta }})$ on $\mathbf{U}_{\widehat{\alpha }\widehat{\beta }}.$ We conclude
that the family $\left[ f(z_{\widehat{\alpha }})\right] _{\widehat{\alpha }%
\in I}$ of local sections $\mathbf{z}$ of $\mathbf{E},$ such that $\mathbf{z}%
_{|U_{\widehat{\alpha }}}=f\circ $ $\mathbf{z}_{\widehat{\alpha }}$ being
distinguished in N--adapted sections $\mathbf{z}_{|U_{\widehat{i}}}=f\circ $
$\mathbf{z}_{\widehat{i}}$ and $\mathbf{z}_{|U_{\widehat{a}}}=f\circ $ $%
\mathbf{z}_{\widehat{a}}.\blacksquare $
\end{proof}

For a chain $z_{\widehat{\alpha }\widehat{\beta }}=$ $z_{\widehat{\alpha }%
}-q_{\widehat{\alpha }\widehat{\beta }}(z_{\widehat{\beta }}),$ we can
construct a 2--cocycle%
\begin{equation*}
z_{\widehat{\beta }\widehat{\gamma }}-z_{\widehat{\alpha }\widehat{\gamma }%
}+z_{\widehat{\alpha }\widehat{\beta }}=z_{\widehat{\alpha }\widehat{\beta }%
\widehat{\gamma }}.
\end{equation*}%
Nevertheless, even there are a chain $q_{\widehat{\alpha }\widehat{\beta }}$
and a global section $\mathbf{z}=\left( \mathbf{z}_{\widehat{\alpha }%
}\right) _{\widehat{\alpha }\in I}$ such that $z_{\widehat{\alpha }}=q_{%
\widehat{\alpha }\widehat{\beta }}(z_{\widehat{\beta }})$ and $f(z_{\widehat{%
\alpha }})=\mathbf{z}_{|U_{\widehat{\alpha }}}$ it may does not exist a
global section for another N--adapted chain. One has to work with the
d--vector space $\mathbf{Z}$ of formal global sections of the N--anholonomic
vector gerbe $\mathbf{C}_{NQ}.$ The d--vector space is defined by generators
$\left[ \mathbf{z}\right] $ where $\mathbf{z}$ is an element of a set of
global sections
\begin{equation*}
\mathbf{Z}\left( q_{\widehat{\alpha }\widehat{\beta }}\right) =\left[
Z\left( q_{\widehat{i}\widehat{j}}\right) ,Z\left( q_{\widehat{a}\widehat{b}%
}\right) \right] .
\end{equation*}%
We can consider that any element of the space $\mathbf{Z}$ is defined by a
formal finite sum of global sections.

\subsubsection{Lifts and d--vector gerbes}

The constructions from the previous section were derived for vector bundles on
N--anholonomic manifolds. But such a nonholomic manifold in its turn has a
local fibered structure resulting in definition of a nonholonomic gerbe $%
\mathbf{C}_{Nd}$ as a d--vectorial gerbe. We denote by $Q^{m}$ the typical
fiber which can be associated to a N--anholonomic manifold $\mathbf{V}$ of
dimension $n+m,$ from the map $\pi :V^{n+m}\rightarrow V^{n},$ see
Definition \ref{dnc}. We can also associate a principal bundle $Gl(Q^{m})$
supposing that this bundle has a reduction $\mathbf{V}_{K}$ for a subgroup $%
K\subset Gl(Q^{m})$ with a central extension of type (\ref{ext1}). For such
an extension, we define a d--vector gerbe $\mathbf{C}_{dH}$ on $h\mathbf{V}$
when for each open set $U\subset h\mathbf{V}$ the objects of $\mathbf{C}%
_{dH}(U)$ are $G$--principal bundles over $U$ when the quotient by $H$ is
the restriction of $\mathbf{V}_{K}$ to $U.$

For the projection $\pi :G\rightarrow K$ and (supposed to exist)
representation $r:G\rightarrow Gl(Q^{\prime m})$ and surjection $f:Q^{\prime
m}\rightarrow Q^{m}$ one can be defined a commutative Diagram 2,

\begin{figure}[tbph]
\begin{center}
\begin{picture}(140,80)\setlength{\unitlength}{1pt}%
\thinlines \put(12,60){${Q^{\prime }}^{m}$}%
\put(100,60){${Q^{\prime }}^m$}%
\put(12,0){$Q^m$}%
\put(100,0){${Q}^m$}%
\put(24,3){ \vector(1,0){65}}%
\put(16,54){\vector(0,-1){40}} %
\put(105,54) {\vector(0,-1){40}}%
\put(8,35){$f$}%
\put(50,12){$\underline{\pi (s)}$}%
\put(50,71){$\underline{r(s)}$}%
\put(108,35){$f$}%
\put(26,62){\vector(1,0){65}}%
\end{picture}
\end{center}
\caption{{Diagram 2}}
\end{figure}
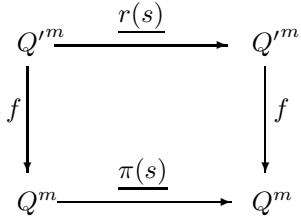
%
%
By such a Diagram, it is defined a d--vectorial gerbe $\mathbf{C}%
_{gH,Q^{\prime }}$ on $h\mathbf{V}$ when an object of $\mathbf{C}(U)$ is
parametrized $e_{U}\propto r$ where $e_{U}$ is an object of $\mathbf{C}%
_{gH}(U).$ The constructions are adapted to the N--connection structure on $U%
\mathbf{.}$ We note that the objects and regions defined with respect to
h--subspaces are not boldfaced as those considered for N--anholonomic vector
bundles. Stating $\left( U_{\widetilde{i}}\right) _{\widetilde{i}\in I}$ as
a trivialization of $h\mathbf{V,}$ with transition functions $q_{\widetilde{i%
}\widetilde{j}}^{\prime },$ we can define the maps $q_{\widetilde{i}%
\widetilde{i}}:U_{\widetilde{i}}\cap U_{\widetilde{i}}\rightarrow G$ over $%
q_{\widetilde{i}\widetilde{j}}^{\prime }.$ This means that $\mathbf{C}%
_{gH,Q^{\prime }}$ is defined by $r(q_{\widetilde{i}\widetilde{j}}).$

There is also is a natural scalar product (a particular case of that for
N--anholonomic gerbers) in our case defined by d--vector gerbes. We can
consider such a scalar product just for the Clifford N--gerbe $Cl(h\mathbf{V}%
)$ following the fact that the group $Spin$ is compact and its action on $Cl(%
\mathbb{R}^{n})$ preserves a scalar product. We conclude that this scalar
product exists for any object $e_{U}$ of $Cl(U)$ and that a d--metric (\ref%
{metr}) states a splitting $<,>_{e_{U}}=<,>_{he_{U}}+<,>_{ve_{U}}.$ There
are some alternatives: There is a family of Riemannian d--metrics on the
d--vector gerbe $Cl(h\mathbf{V})$ but this is not adapted to the
N--connection structure. \ One has to apply the concept of d--connection in
order to define N--adapted objects. If the d--metric structure is not
prescribed, we can introduce a scalar product structure defined by the
N--connection when $(g_{ij},h_{ab})$ in (\ref{metr}) are taken to be some
Euclidean ones but $N_{i}^{a}(\mathbf{u})$ are the coefficients for a
nontrivial N--connection.

\begin{definition}
A d--metric \ (it is connected to a N--anholonomic Riemann--Cartan
structure) on a d--vector gerbe $\mathbf{C}_{Nd}$ is given by a
distinguished scalar product $<,>_{e_{U}}=(<,>_{he_{U}},<,>_{ve_{U}})$ on
the d--vector bundle $\mathbf{e}_{U},$ defined for every object of $\mathbf{C%
}_{Nd}$ and preserved by morphisms of such objects.
\end{definition}

Following this Definition, for the d--vector gerbes, one holds the
Proposition \ref{pnvg} and related results.

\subsection{Pre--Hibertian and d--connection structures}

For simplicity, hereafter we shall work only with N--anhlonomic manifolds.
We emphasize that the constructions can be extended to vector bundles $%
\mathbf{E}$ on such a nonhonomic manifold $\mathbf{V}.$ The proofs will be
omitted if they are similar to those given for holonomic manifolds and
vector spaces \cite{aris} but (in our case) adapted to the splitting defined
by the N--connection structure. We shall point out the nonholonomic character
of the constructions. Such computations may be performed directly by
applying ''boldfaced'' objects.

\subsubsection{Distinguished pre--Hilbertian and scalar structures}

Let us consider the two elements $\ _{1}z$ $\ $\ and $\ _{2}z$ of the
d--vector space $\mathbf{Z}\left( q_{\widehat{\alpha }\widehat{\beta }%
}\right) =\left[ Z\left( q_{\widehat{i}\widehat{j}}\right) ,Z\left( q_{%
\widehat{a}\widehat{b}}\right) \right] $ (we use left low labels which are
not indices running values). For a partition of unity $(\mathbf{U}_{\widehat{%
\alpha }^{\prime }}^{\prime },f_{\widehat{\alpha }^{\prime }})_{\widehat{%
\alpha }^{\prime }\in I^{\prime }}$ subordinated to $\left( \mathbf{U}_{%
\widehat{\alpha }}\right) _{\widehat{\alpha }\in I}.$ \footnote{%
this means that for each $\widehat{\alpha }^{\prime }$ there is an $\widehat{%
\alpha }(\widehat{\alpha }^{\prime })$ such that $\mathbf{U}_{\widehat{%
\alpha }^{\prime }}^{\prime }$ is a subset of $\mathbf{U}_{\widehat{\alpha }(%
\widehat{\alpha }^{\prime })}$} Since the support of $f_{\widehat{\alpha }%
^{\prime }}$ is a couple of compact subsets of $\mathbf{U}_{\widehat{\alpha }%
^{\prime }}^{\prime }$ distinguished by the N--connection structure, we can
consider restrictions of $\ _{1}z_{\widehat{\alpha }(\widehat{\alpha }%
^{\prime })}$ $\ $\ and $\ _{2}z_{\widehat{\alpha }(\widehat{\alpha }%
^{\prime })}$ to $\mathbf{U}_{\widehat{\alpha }^{\prime }}^{\prime }$
denoted respectively $\ _{1}z_{\widehat{\alpha }\widehat{\alpha }(\widehat{%
\alpha }^{\prime })}$ $\ $\ and $\ _{2}z_{\widehat{\alpha }\widehat{\alpha }(%
\widehat{\alpha }^{\prime })}.$ We can calculate the value $\int <\ _{1}z_{%
\widehat{\alpha }\widehat{\alpha }(\widehat{\alpha }^{\prime })},\ _{2}z_{%
\widehat{\alpha }\widehat{\alpha }(\widehat{\alpha }^{\prime })}>$ which is
invariant for any partition. This proofs

\begin{proposition}
The scalar product
\begin{eqnarray}
&<&\ _{1}z,\ _{2}z>\ =\sum_{\widehat{\alpha }^{\prime }}\int <f_{\widehat{%
\alpha }^{\prime }}\ _{1}z_{\widehat{\alpha }\widehat{\alpha }(\widehat{%
\alpha }^{\prime })},f_{\widehat{\alpha }^{\prime }}\ _{2}z_{\widehat{\alpha
}\widehat{\alpha }(\widehat{\alpha }^{\prime })}>  \label{phds} \\
&=&\sum_{\widehat{i}^{\prime }}\int <f_{\widehat{i}^{\prime }}\ _{1}z_{%
\widehat{i}\widehat{i}(\widehat{i}^{\prime })},f_{\widehat{i}^{\prime }}\
_{2}z_{\widehat{i}\widehat{i}(\widehat{i}^{\prime })}>+\sum_{\widehat{a}%
^{\prime }}\int <f_{\widehat{a}^{\prime }}\ _{1}z_{\widehat{a}\widehat{a}(%
\widehat{a}^{\prime })},f_{\widehat{a}^{\prime }}\ _{2}z_{\widehat{a}%
\widehat{a}(\widehat{a}^{\prime })}>  \notag
\end{eqnarray}%
defines a pre--Hilbert d--structure of $\left( \mathbf{Z}\left( q_{\widehat{%
\alpha }\widehat{\beta }}\right) ,<,>\right) .$ \ \
\end{proposition}

For two formal global sections of d--vector gerbe $\mathbf{C}_{Nd},$ we can
write
\begin{equation*}
\ _{1}z=\left[ z_{\beta _{1}}\right] +...+\left[ z_{\beta _{p}}\right] \ %
\mbox{ and }\ _{2}z=\left[ z_{\gamma _{1}}\right] +...+\left[ z_{\gamma _{p}}%
\right]
\end{equation*}%
where $z_{\beta _{p}}$ and $z_{\gamma _{p}}$ are global sections. The scalar
product on the space $\mathbf{Z,}$ of formal global sections of the
d--vector gerbe $\mathbf{C}_{Nd},$ can be defined by the rule%
\begin{equation*}
<\left[ \ _{1}z\right] ,\left[ \ _{2}z\right] >\ =\ <\ _{1}z,\ _{2}z>_{%
\mathbf{Z}\left( q_{\widehat{\alpha }\widehat{\beta }}\right) }
\end{equation*}%
if $\ _{1}z$ and $\ _{2}z$ are elements of the same set of global sections $%
\mathbf{Z}\left( q_{\widehat{\alpha }\widehat{\beta }}\right) $ and
\begin{equation*}
<\left[ \ _{1}z\right] ,\left[ \ _{2}z\right] >\ =0
\end{equation*}%
for the elements belonging to different sets of such global sections.

\begin{proposition}
Any element $\mathbf{z}_{\widehat{\alpha }}$ of the Hilbert completion $%
L^{2}(\mathbf{Z}\left( q_{\widehat{\alpha }\widehat{\beta }}\right) )$ of
the pre--Hilbert d--structure (\ref{phds}) is a N--adapted family of $L^{2}$
sections $\mathbf{z}_{\widehat{\alpha }}$ of $\mathbf{e}_{\widehat{\alpha }}$
such that $\mathbf{z}_{\widehat{\alpha }}=q_{\widehat{\alpha }\widehat{\beta
}}(z_{\widehat{\beta }}).$
\end{proposition}

\begin{proof}
The proof is similar to that for the Proposition 7 in Ref. \cite{aris} and
follows defining a corresponding Cauchy sequence $\left( z^{\widehat{\alpha }%
}\right) _{\widehat{\alpha }\in \mathbb{N}}$ of \ \newline
$\left( \mathbf{Z}\left( q_{\widehat{\alpha }\widehat{\beta }}\right)
,<,>\right) .$ For nonholonomic configurations, one uses d--metric
structures which can be Riemannian or Riemann--Cartan ones depending on the
type of linear connection is considered, a not N--adapted, or N--adapted one.%
$\blacksquare $
\end{proof}

We can consider morphisms between d--objects commuting with Laplacian $%
\bigtriangleup ^{s}$ and define a pre--Hilbertian structure defined by
\begin{equation*}
<\ _{1}z,\ _{2}z>\ =\int <\bigtriangleup ^{s}(\ _{1}z),\ _{2}z>,
\end{equation*}%
where $\bigtriangleup ^{s}(\ _{1}z)_{\widehat{\alpha }}=\bigtriangleup
^{s}(\ _{1}z_{\widehat{\alpha }}).$ There is a canonical N--adapted
Laplacian structure \cite{vncl,vfs,vv} defined on nonholonomic spaces by
using the canonical d--connection structure, see Proposition \ref{pcdc} in
Appendix. We denote by $H_{s}(\mathbf{Z}\left( q_{\widehat{\alpha }\widehat{%
\beta }}\right) )$ the Hilbert completion of the pre--Hilbert space
constructed by using $\bigtriangleup ^{s}.$

\subsubsection{D--connections on N--anholonomic gerbes and characteristic
classes}

The canonical d--connection structure gives rise \ to a such connection on
each d--object $\mathbf{e}_{U}$ of $\mathbf{C}(\mathbf{U})$ and defined a
family of d--connections inducing such a structure on the N--anholonomic
gerbe $\mathbf{C}.$ We consider an open covering $\left( \mathbf{U}_{%
\widehat{\alpha }}\right) _{\widehat{\alpha }\in I}$ of $\mathbf{V}$ and
d--objects $\mathbf{C}(\mathbf{U}_{\widehat{\alpha }})$ as trivial bundles
with $m$--dimensional fibers. The d--connection $\mathbf{\Gamma }_{\widehat{%
\alpha }}$ of a d--object $\mathbf{e}_{\widehat{\alpha }}$ of $\mathbf{C}(%
\mathbf{e}_{\widehat{\alpha }})$ is defined by a 1--form with coefficients (%
\ref{candcon}) on $T$ $\mathbf{U}_{\widehat{\alpha }}.$ The curvature of
this d--connection is $\mathcal{R}_{\widehat{\alpha }},$ see (\ref{curv}).

Having defined the curvature of the N--anholonomic gerbe, it \ is possible
to compute the $2k$ Chern class of $\mathbf{e}_{\widehat{\alpha }},$%
\begin{equation*}
c_{2k}^{\widehat{\alpha }}\sim Tr\left[ \left( \frac{i}{2\pi }\mathcal{R}_{%
\widehat{\alpha }}\right) ^{k}\right] ,
\end{equation*}%
where $Tr$ denotes the trace operation, which is invariant for transforms $%
\mathbf{e}_{\widehat{\alpha }}\rightarrow \mathbf{e}_{\widehat{\alpha }%
}^{\prime }.$ In a more general case, we can compute the sum%
\begin{equation*}
c(\mathbf{C})=c_{1}(\mathbf{V})+...+c_{2}(\mathbf{V})
\end{equation*}%
for the total Chern form of an N--anholonomic gerbe. This form define locally
the total Chern character
\begin{equation}
ch(\mathbf{C})_{\mathbf{U}_{\widehat{\alpha }}}=Tr\left[ \exp \left( \frac{i%
}{2\pi }\mathcal{R}\right) \right] .  \label{chkh1}
\end{equation}

It should be noted that the formula (\ref{chkh1}) is defined by a d--metric (%
\ref{metr}) and its canonical d--connection (\ref{cdc}) which correspond
respectively to the Riamannian metric and the Levi--Civita connection. The
notion of connection is not well--defined for general vector gerbes but the
existence of Riemannian structures gives a such possibility. In the case of
N--anholonomic frame, even a d--metric structure is not stated, we can derive
a canonical d--connection configuration by considering a formal d--metric
with $g_{ij}$ and $h_{ab}$ taking diagonal Euclidean values and computing a
curvature tensor $\mathcal{R}^{[N]}$ by contracting the N--connection
coefficients $N_{i}^{a}$ and theirs derivatives. As a matter of principle,
we can take the N--connection curvature $\Omega $ (\ref{ncurv}) instead of $%
\mathcal{R}^{[N]}$ but in this case we shall deal with metric noncommpatible
d--connections. Finally, we not that we need at leas to Chern characters,
one for the d--connection structure \ and another one for the N--connection
structure in order to give a topological characteristic of N--anholonomic
gerbes.

\section{Operators and Symbols on Nonholonomic Ger\-bes}

On N--anholonomic manifolds we deal with geometrical objects distinguished
by a N--connection structure. The aim of this section is to analyze
pseudo--differential operators $\mathbf{D}_{\overline{\alpha }}$ on such 
spaces.

\subsection{Operators on N--anholonomic spaces}

In local form, the geometric constructions adapted to a N--connection are
for open sets of couples $\left( \mathbb{R}^{n},\mathbb{R}^{m}\right) ,$ or
for $\mathbb{R}^{n+m}.$ Let us consider an open set $\mathbf{U}\subset
\mathbb{R}^{n+m}$ and denote by $Z^{r}(\mathbf{U})$ the set of smooth
functions $p(\mathbf{v,u})$ defined on $\mathbf{U}\times \mathbb{R}^{n+m}$
satisfying the conditions that for any compact $\mathbf{U}^{\prime }\subset
\mathbf{U}$ and every multi--indices $\alpha $ and $\beta $ one has
\begin{equation*}
\left\| \mathbf{D}^{\overline{\alpha }}\mathbf{D}^{\overline{\beta }}p(%
\mathbf{v,u})\right\| <C_{\overline{\alpha },\overline{\beta },\mathbf{U}%
^{\prime }}\left( 1+||u||\right) ^{r-|\overline{\alpha }|},
\end{equation*}%
for $C_{\overline{\alpha },\overline{\beta },\mathbf{U}^{\prime }}=const.$

\begin{definition}
A map $\widehat{p}$ of two smooth functions $k(\mathbf{U})$ and $k^{\prime }(%
\mathbf{U})$ with compact support defined on $\mathbf{U,}$ $\widehat{p}:$ $k(%
\mathbf{U})$ $\rightarrow $ $k^{\prime }(\mathbf{U}),$ such that locally
\begin{equation*}
\widehat{p}(f)=\int p(\mathbf{v,u})\widehat{f}(\mathbf{u})e^{i<\mathbf{v,u}%
>}\delta \mathbf{u,}
\end{equation*}%
where $\widehat{f}$ is the Fourier transform of function $f,$ is called to
be a pseudo--differential distinguished operator, in brief \textit{pdd}%
--operator.
\end{definition}

Now we extend the concept of \textit{pdd}--operator for a N--anholonomic
manifold $\mathbf{V}$ endowed with d--metric structure (\ref{metr}). In this
case, $k(\mathbf{U})$ and $k^{\prime }(\mathbf{U})$ are smooth sections,
with compact support, of $\mathbf{V}$ provided with local fibered structure.
A map $\widehat{p}$ is defined for a covering family $\left( \mathbf{U}_{%
\widehat{\alpha }}\right) _{\widehat{\alpha }\in I}$ satisfying the
conditions:

\begin{enumerate}
\item Any restriction of $\mathbf{V}$ to $\mathbf{U}_{\widehat{\alpha }}$ is
trivial.

\item The map $\widehat{p}_{\widehat{\alpha }}:k(\mathbf{U}_{\widehat{\alpha
}}\mathbf{\times }Q^{m})$ $\rightarrow $ $k^{\prime }(\mathbf{U}_{\widehat{%
\alpha }}^{\prime }\mathbf{\times }Q^{m}),$ where the vector space $Q^{m}$
is isomorphic to $v\mathbf{V,}$ $dim(v\mathbf{V})=m,$ defines the
restriction of $\widehat{p}$ to $\mathbf{U}_{\widehat{\alpha }}.$ There is a
horizontal component of the map, $\widehat{p}_{\widehat{i}}:k(U_{\widehat{i}}%
\mathbf{\times }Q^{m})$ $\rightarrow $ $k^{\prime }(U_{\widehat{i}}^{\prime }%
\mathbf{\times }Q^{m})$

\item For any section $z^{\prime }$ over $h\mathbf{U}_{\widehat{\alpha }}=U_{%
\widehat{i}}$ and $z=(z_{1},...z_{m})=\phi _{a}(z^{\prime }),$ we can define%
\begin{equation*}
t_{b}=\sum\limits_{a=1}^{a=m}\int p_{ab}(x^{i},v^{k})\widehat{z}%
_{a}(v^{l})e^{i<x,v>}\delta v^{j}
\end{equation*}%
and $\widehat{p}_{a}(z^{\prime })=\psi _{a}^{-1}(t_{b}),$ where the carts $%
\phi _{a}$ and $\psi _{a}$ are such that
\begin{equation*}
\phi _{a}(U_{\widehat{i}}\mathbf{\times }Q^{m})=\psi _{a}(U_{\widehat{i}}%
\mathbf{\times }Q^{m})\simeq h\mathbf{U\times }\mathbb{R}^{m}
\end{equation*}%
and the map $\widehat{p}_{a}$ is defined by a matrix $p_{ab}$ defining an
operator of degree $r.$

\item The values $t_{b},z_{a},\phi _{a},\psi _{a}$ and $p_{ab}$ can be
extended to corresponding distinguished objects
\begin{eqnarray*}
t_{b} &\rightarrow &t_{\alpha }=(t_{i},t_{b}),z_{a}\rightarrow z_{\alpha
}=(z_{i},z_{a}), \\
\phi _{a} &\rightarrow &\phi _{\alpha }=(\phi _{i},\phi _{a}),\psi
_{a}\rightarrow \psi _{\alpha }=(\psi _{i},\psi _{a}),p_{ab}\rightarrow
p_{\alpha \beta }.
\end{eqnarray*}
\end{enumerate}

\begin{definition}
\label{dpddnm}A map $\widehat{p}$ satisfying the conditions 1-4 defines a
\textit{pdd}--operator on N--anholonomic manifold $\mathbf{V}$ provided with
d--metric structure (\ref{metr}). For Euclidean values for h- and
v--components of d--metric, with respect to N--adapted frames, one gets a
\textit{pdd}--operator generated by the N--connection structure.
\end{definition}

The Definition \ref{dpddnm} can be similarly formulated for N--anholonomic
vector bundles $\mathbf{E}\rightarrow \mathbf{V}.$ We denote by $\
^{loc}H_{s}(\mathbf{V,E}),$ with $s$ being a positive integer, the space of
distributions sections $\underline{\mathbf{u}}$ of $\mathbf{E}$ such that $%
\mathbf{D}(\underline{\mathbf{u}})$ is a $\ ^{loc}L^{2}$ section, where $%
\mathbf{D}$ is any differential d--operator of order less than $s.$ The
subset of elements of $\ ^{loc}H_{s}(\mathbf{V,E})$ with compact support is
written $\ ^{comp}H_{s}(\mathbf{V,E}).$ The space $\ ^{loc}H_{-s}(\mathbf{V,E%
})$ is defined to be the dual space of $\ ^{comp}H_{s}(\mathbf{V,E})$ and
the space $\ ^{comp}H_{-s}(\mathbf{V,E})$ is defined to be the dual space of
$\ ^{loc}H_{s}(\mathbf{V,E}).$

\begin{definition}
The Sobolev canonical d--space $H_{s}$ is an Hilbert space provided with the
norm%
\begin{equation*}
\left\| \int <\widehat{\bigtriangleup }^{s}\underline{\mathbf{u}},\underline{%
\mathbf{u}}>\right\| ^{1/2}
\end{equation*}
defined by the Laplace operator $\widehat{\bigtriangleup }^{s}\doteqdot
\widehat{\mathbf{D}}_{\alpha }\widehat{\mathbf{D}}^{\alpha }$ of the
canonical d--connection structure (\ref{cdc}). Every d--operator $\widehat{p}
$ of order less than $r$ can be extended to a continuous morphism $%
H_{s}\rightarrow H_{s-r}.$
\end{definition}

We can generalize the last two Definitions for N--anholonomic gerbes:

\begin{definition}
A d--operator $\mathbf{D}$ of degree $r$ on N--anholonomic gerbe $\mathbf{C}
$ provided with d--metric and canonical d--connection structures is defined
by a family of operators $\mathbf{D}_{e}$ of degree $r$ defined on an object
$\mathbf{e}$ of the category $\mathbf{C}(\mathbf{U})$ when for each morphism $%
\varphi :\mathbf{e}\rightarrow f$ one holds $\mathbf{D}_{f}\varphi
^{\#}=\varphi ^{\#}\mathbf{D}_{e}.$
\end{definition}

In this Definition the map $\varphi ^{\#}$ transforms a section $z$ to $%
\varphi (z)$ and it is supposed that $\mathbf{D}_{e}$ is invariant under
N--adapted authomorphisms of $\mathbf{e}.$ It is also assumed to be a
continuous operator as a map
\begin{equation*}
\mathbf{D}_{e}:\ ^{comp}H_{s}(\mathbf{U,e})\rightarrow \ ^{loc}H_{s-r}(%
\mathbf{U,e}).
\end{equation*}%
For a global distributional section $z$ as an element of $H_{s}(\mathbf{Z}%
_{q_{\alpha \beta }}),$ we can write $q_{\widehat{\alpha }\widehat{\beta }%
}\left( \mathbf{D}_{e_{\widehat{\beta }}^{\widehat{\alpha }}}(e_{\widehat{%
\beta }}^{\widehat{\alpha }})\right) =\left( \mathbf{D}_{e_{\widehat{\alpha }%
}^{\widehat{\beta }}}(e_{\widehat{\alpha }}^{\widehat{\beta }})\right) .$
This proves

\begin{proposition}
Any d--operator $\mathbf{D}$ of degree $r$ defined on an N--anholono\-mic
gerbe $\mathbf{C}$ provided with d--metric and canonical d--connection
structures induces two maps%
\begin{equation*}
\mathbf{D}_{\mathbf{Z}\left( q_{\widehat{\alpha }\widehat{\beta }}\right)
}:H_{s}\left( \mathbf{Z}\left( q_{\widehat{\alpha }\widehat{\beta }}\right)
\right) \rightarrow H_{s-r}\left( \mathbf{Z}\left( q_{\widehat{\alpha }%
\widehat{\beta }}\right) \right) \mbox{ and }\mathbf{D}_{\mathbf{Z}%
}:H_{s}\left( \mathbf{Z}\right) \rightarrow H_{s-r}\left( \mathbf{Z}\right) .
\end{equation*}
\end{proposition}

In this paper, we shall consider only \textit{pdd}--operators preserving $%
C^{\infty }$ sections.

\subsection{The symbols of nonholonomic operators}

Let us consider a \textit{pdd}--operator $\widehat{p}_{\widehat{\alpha }%
}(f)=\int p_{\widehat{\alpha }}(\mathbf{v,u})\widehat{f}(\mathbf{u})e^{i<%
\mathbf{v,u}>}\delta \mathbf{u}$ defined for a restriction of $\widehat{p}$
to $\mathbf{U}_{\widehat{\alpha }}$ from a covering $\left( \mathbf{U}_{%
\widehat{\alpha }}\right) _{\widehat{\alpha }\in I}$ of an open $\mathbf{%
U\subset V.}$

\begin{definition}
The operator $\widehat{p}_{\widehat{\alpha }}$ is of degree $r$ with the
symbol $\sigma (\mathbf{p})$ if there exist the limit $\sigma (\mathbf{p}%
_{\mid \mathbf{U}_{\widehat{\alpha }}})=\lim_{\lambda \rightarrow \infty
}\left( p_{\widehat{\alpha }}(\mathbf{v,}\lambda \mathbf{u})/\lambda
^{r}\right) .$
\end{definition}

This definition can be extended for a sphere bundle $S\mathbf{V}$ of the
cotangent bundle $T^{\ast }\mathbf{V}$ of $\mathbf{V}$ and for $\pi ^{\ast }%
\mathbf{E}$ being the pull--back of the vector bundle $\mathbf{E}$ on $%
\mathbf{V}$ to $T^{\ast }\mathbf{V.}$ The symbols defined by the matrix $%
p_{\alpha \beta }$ define a map $\sigma :\pi ^{\ast }\mathbf{E\rightarrow }%
\pi ^{\ast }\mathbf{E.}$ This map also induces a map $\sigma _{S}:\pi
_{S}^{\ast }\mathbf{E\rightarrow }\pi _{S}^{\ast }\mathbf{E}$ if we consider
the projection $\pi _{S}:S\mathbf{V\rightarrow V.}$

Now we analyze the symbols of operators on a N--anholonomic vector gerbe $%
\mathbf{C}_{NQ}$ defined on $\mathbf{V}$ endowed with the operator $D$ of
degree $r.$ For each object $\mathbf{e}$ of $\mathbf{C}(\mathbf{U}),$ it is
possible to pull back the bundle $\mathbf{e}$ by the projection map $\pi
_{SU}:S\mathbf{U\rightarrow U}$ to a bundle $\pi _{S\mathbf{U}}^{\ast }e$
over $S\mathbf{U.}$ This nonholonomic bundle is the restriction of the
co--sphere bundle defined by a fixed d--metric on $T^{\ast }\mathbf{V.}$ We
can define a category consisting from the family $\mathbf{C}_{S}(\mathbf{U}%
) $ with elements $\pi _{SU}^{\ast }e$ and baps of such elements induced by
maps between elements of $\mathbf{C}(\mathbf{U}).$ In result, we can
consider that the distinguished by N--connection map $\mathbf{U}\rightarrow
C_{S}(\mathbf{U})$ is a N--anholonomic gerbe with the same band as for $%
\mathbf{C}. $ For an object $e,$ it is possible to define the symbol $\sigma
_{D_{e}}:\pi _{S\mathbf{U}}^{\ast }e\rightarrow \pi _{S\mathbf{U}}^{\ast }e.$

\begin{proposition}
For any sequence $f_{k}$ of elements of $H_{s}\left( \mathbf{Z}\left( q_{%
\widehat{\alpha }\widehat{\beta }}\right) \right) $ and a constant $f_{[0]}$
such that $\left\| f_{k}\right\| _{s}<f_{[0]},$ there is a subsequence $%
f_{k^{\prime }}$ converging in $H_{s^{\prime }}$ for any $s>s^{\prime }.$
\end{proposition}

The proof follows from the so--called Relich Lemma (Proposition 5) in Ref. %
\cite{aris}: we have only to consider it both for the so--called h-- and
v--subspaces. For simplicity, in this subsection, we outline only some basic
properties of d--operators for N--anholonomic gerbes which are distinguished
by the N--connection structure: First, the space $Op(\mathbf{C})$ of
continuous linear N--adapted maps of $H_{s}\left( \mathbf{Z}\left( q_{%
\widehat{\alpha }\widehat{\beta }}\right) \right) $ is a Banach space.
Secondly, the last Preposition states the possibility to define $O^{r}$ the
completion of the pseudo--differential operators in $OP(\mathbf{C})$ of
order $r$ and to extend the symbol $\sigma $ to this completion. Finally,
the kernel of such an extension of the symbol to $O^{r}$ contains only
compact operators.

\section{$ K$--Theory and the N--Adapted Index}

This section is devoted to $K$--theory groups $K_{0}$ and $K_{1}$ associated
to symbols of d--operators on N--anholonomic gerbes as elements of $%
K_{0}(T^{\ast }\mathbf{V}).$

\subsection{$K$--theory groups $K_{0}$ and $K_{1}$ and N--anholonomic spaces}

We show how some basic results from $K$--theory (see, for instance, Ref. %
\cite{karoubi}) can be applied for nonholonomic manifolds.

\subsubsection{Basic definitions}

Let us denote by $A_{n}$ the vector space of $n\times n$ complex matrices,
consider natural injections $A_{n}\rightarrow A_{n^{\prime }}$ for $n\leq
n^{\prime }$ and denote by $A_{\infty }$ the inductive limit of the vector
space $A_{n},n\in \mathbb{N}.$ For a ring $B$ and two idempotents $a^{\prime
}$ and $b^{\prime }$ of $B_{\infty }=B\otimes A_{\infty },$ one says that $%
a\sim b$ if and only if there exists elements $a^{\prime },b^{\prime }\in
A_{\infty }$ such that $a=a^{\prime }b^{\prime }$ and $b=b^{\prime
}a^{\prime }.$ Let us denote by $[a]$ the class of $a$ and by $%
Idem(B_{\infty })$ the set of equivalence classes. Representing,
respectively, $[a]$ and $[b]$ by elements of $B$ $\otimes A_{n}$ and $B$ $%
\otimes A_{n^{\prime }},$ we can define an idempotent of $B$ $\otimes
A_{n+n^{\prime }}$ represented by $[a+b]=\left(
\begin{array}{cc}
a & 0 \\
0 & b%
\end{array}%
\right) .$ The semi--group $Idem(B_{\infty })$ provided with the operation $%
[a]+[b]=[a+b]$ is denoted by $K_{0}(B).$

One can extend the construction for a compact N--manifold $\mathbf{V}$ and a
the set of complex valued functions $\mathbb{C(}\mathbf{V}\mathbb{)}$ on $%
\mathbf{V.}$ For compact manifolds, it is possible to consider a trivial
bundle isomorphic to $\mathbf{V\times }\mathbb{C}^{r}$ and identify a vector
bundle over $\mathbf{V}$ to an idempotent of $\mathbb{C(}\mathbf{U}\mathbb{%
)\otimes }A_{r},$ for $\mathbf{U\subset V,}$ which is also an idempotent of $%
\mathbb{C}(\mathbf{U})_{\infty }.$ This allows us to identify $K_{0}\mathbb{(%
}\mathbf{V}\mathbb{)}$ to $K_{0}\mathbb{(\mathbb{C}}(\mathbf{U})).$ Such a
group for N--anholonomc manifold is a distinguished one, i.e. d--groups,
into two different components, respectively for the h--subspace and the
v--subspaces of $\mathbf{V.}$

Now we define the $K_{1}$ group: Let $Al_{r}(B)$ is the group of invertible
elements contained in the matrix group $A_{r}(B)$. For $r^{\prime }\leq r,$
there is a canonical inclusion map $Al_{r^{\prime }}(B)\rightarrow
Al_{r}(B). $ The group $Al_{\infty }(B)$ denotes the inductive limit of the
groups $Al_{r}(B)$ and $Al_{\infty }(B)_{con}$ is the respective connected
component. In result, the group $K_{1}(B)$ is the quotient $Al_{\infty
}(B)/Al_{\infty }(B)_{con}.$ For a compact N--anholonomic manifold $\mathbf{%
V,}$ one defines $K_{1}\mathbb{(}\mathbf{V}\mathbb{)}$ by $K_{1}\mathbb{(%
\mathbb{C}}(\mathbf{U})).$

\subsubsection{N--anholonomic elliptic operators and indices}

We consider a N--anholonomic gerbe $\mathbf{C}$ on manifold $\mathbf{V}$ and
an elliptic operator $D$ of degree $r$ on $\mathbf{C,}$ inducing a
distinguished morphism
\begin{equation*}
D:L^{2}\left( \mathbf{Z}\left( q_{\widehat{\alpha }\widehat{\beta }}\right)
\right) \rightarrow H_{2-r}\left( \mathbf{Z}\left( q_{\widehat{\alpha }%
\widehat{\beta }}\right) \right) .
\end{equation*}
Let $(1-\bigtriangleup )^{-r^{\prime }}$ be an operator of degree $-r,$ for
instance, we can take $\bigtriangleup $ to be the Laplace operator $\widehat{%
\bigtriangleup }^{s}\doteqdot \widehat{\mathbf{D}}_{\alpha }\widehat{\mathbf{%
D}}^{\alpha }$ of the canonical d--connection structure (\ref{cdc}). It is
possible to define the symbol $\sigma (D)=(1-\bigtriangleup )^{-r^{\prime
}}D $ which is an operator with image in \ $\mathbf{Z}\left( q_{\widehat{%
\alpha }\widehat{\beta }}\right) $ being an invertible morphism.

For an exact sequence
\begin{equation*}
0\rightarrow B_{1}\rightarrow B_{2}\rightarrow B_{3}\rightarrow 0
\end{equation*}%
of $C^{\ast }$--algebras, one has the following exact sequence in $K$--theory%
\begin{equation*}
K_{1}\left( B_{1}\right) \rightarrow K_{1}\left( B_{2}\right) \rightarrow
K_{1}\left( B_{3}\right) \rightarrow K_{0}\left( B_{1}\right) \rightarrow
K_{0}\left( B_{2}\right) \rightarrow K_{0}\left( B_{3}\right) .
\end{equation*}%
Let us consider $O(\mathcal{H}),$ the space of continuous operators on an
Hilbert space $\mathcal{H},$ and denote by $\mathcal{K}$ the subspace of
compact continuous operators. There is the following exact sequence
\begin{equation*}
0\rightarrow \mathcal{K\rightarrow }O(\mathcal{H})\rightarrow O(\mathcal{H})/%
\mathcal{K}=\mathbf{Ca}\rightarrow 0
\end{equation*}%
when $K_{0}\left( \mathcal{K}\right) =\mathbb{Z}.$ Now, it is possible to
introduce the class $\left[ \sigma (D^{\prime })\right] ,$ for $D^{\prime
}=(1-\bigtriangleup )^{-r^{\prime }}D,$ of $K_{1}(\mathbf{Ca}).$

\begin{definition}
The image of $\left[ \sigma (D^{\prime })\right] $ in $K_{0}\left( \mathcal{K%
}\right) $ depends only on the symbol of operator $D$ and define the index
of this operator.
\end{definition}

We consider finite covering families $\mathbf{U}_{\widehat{\alpha }}$ of $%
\mathbf{V}$ when $\mathbf{C(U}_{\widehat{\alpha }}\mathbf{)}$ are trivial
bundles. One holds

\begin{proposition}
One exists a class $\left[ \sigma _{D}\right] $ in $K_{1}(T^{\ast }\mathbf{V}%
)$ associated to the symbol of an elliptic operator $D$ of degree $r$ on
N--anholonomic gerbe $\mathbf{C}$ on $\mathbf{V.}$
\end{proposition}

\begin{proof}
A similar result is proven in \cite{aris} for the Riemannian gerbes. We do
not repeat those constructions in distinguished form for h- and
v--components but note two important differences: In the N--anholonomic case
there are N--connections, d--metrics and d--connections. In result, one can
follow two ways: to define the class for the canonical d--connection and/or
to derive the class from the N--connection structure and related curvature
of N--connection.
\end{proof}

We denote by $X^{\ast }\mathbf{V}$ (with the fibers isomorphic to the unit
ball defined by the N--connection) the compactification of $T^{\ast }\mathbf{%
V.}$ The sphere N--anholonomic bundle $S^{\ast }\mathbf{V}$ is identified to
$X^{\ast }\mathbf{V/}T^{\ast }\mathbf{V.}$ In result, one can define the
exact sequence
\begin{equation*}
0\rightarrow \mathbf{C(}T\mathbf{^{\ast }\mathbf{V})\rightarrow C(}X\mathbf{%
^{\ast }\mathbf{V)}\rightarrow C(}S\mathbf{^{\ast }\mathbf{V)\rightarrow }}0
\end{equation*}%
resulting to the following exact sequence
\begin{eqnarray}
K_{1}\left[ \mathbf{C(}S\mathbf{^{\ast }\mathbf{V)\otimes }}A_{r}\right]
&\rightarrow &K_{0}\left[ \mathbf{C(}T\mathbf{^{\ast }\mathbf{V})\mathbf{%
\otimes }}A_{r}\right]  \label{seq01} \\
&\mathbf{\rightarrow }&K_{0}\left[ \mathbf{C(}X\mathbf{^{\ast }\mathbf{%
V)\otimes }}A_{r}\right] \mathbf{\rightarrow }K_{1}\left[ \mathbf{C(}S%
\mathbf{^{\ast }\mathbf{V)\otimes }}A_{r}\right] \rightarrow 0.  \notag
\end{eqnarray}%
The last sequence allows us to consider the boundary operator $\delta (\left[
\sigma _{D}^{\prime }\right] )$ as an element of $K_{0}\left[ \mathbf{C(}T%
\mathbf{^{\ast }\mathbf{V})\mathbf{\otimes }}A_{r}\right] $ being isomorphic
to $K_{0}\left[ T\mathbf{^{\ast }\mathbf{V}}\right] $ where
\begin{equation*}
\left[ \sigma _{D}^{\prime }\right] \in K_{1}\left[ \mathbf{C(}S\mathbf{%
^{\ast }\mathbf{V)\otimes }}A_{r}\right] \simeq K_{1}\left[ \mathbf{C(}S%
\mathbf{^{\ast }\mathbf{V)}}\right]
\end{equation*}%
for any N--anholonomic component of the sequence (\ref{seq01}). This
concludes that the index of a d--operator $D$ depends only on the class of
boundary operator $\delta (\left[ \sigma _{D}^{\prime }\right] ).$ For
N--anholonomic configurations, such constructions are possible both the
canonical d--connection and if it is not defined by a d--metric, one can
re--define the constructions just for the N--connection and related metric
compatible and N--adapted linear connection and resulting curvatures.

It should be noted that the class $\left[ \sigma _{D}\right] $ is not
unique. For N--anholonomic spaces, we can define such classes, for instance,
by using the canonical d--connection or following d--metrics and
d--connections derived from the N--connection structure.

\subsection{The index formulas and applications}

The results stated in previous subsections allow us to deduce Atiyah--Singer
type theorems for N--anholonomic gerbes (in general form, for any their
explicit realizations like Lagrange, or Finsler, gerbes and Riemann--Cartan
gerbes provided with N--connection structure). Such theorems my have a
number of applications in modern noncommutative geometry and physics. We
shall consider the topic related to Dirac d--operators and N--anholonomic
gerbes.

\subsubsection{The index formulas for d--operators and gerbes}

The Chern character of the cotangent bundle $T^{\ast }M$ induces a well
known isomorphism $K_{0}(T^{\ast }M)\otimes \mathbb{R}\rightarrow \
^{ev}H_{c}(M,\mathbb{R})$ when for elements $u^{\ast }\subset T^{\ast }M$
and $t\in \mathbb{R}$ one has the map $u^{\ast }\otimes t\rightarrow
tch(u^{\ast }).$ The constructions may be generalized for N--anholonomic
gerbes, see (\ref{chkh1}), with additional possibilities related to the
N--connection and d--connection structures.

We denote by $d^{\ast }Vect(Ind)$ the subspace of $K_{0}(\mathbf{V}^{\ast
}), $ with N--anholonomic manifold $\mathbf{V}^{\ast }$ constructed to have
local coordinates $\ ^{\ast }u^{\alpha }=(x^{i},\ ^{\ast }y_{a}),$ with $\
^{\ast }y_{a}$ being dual to $y^{a},$ where $u^{\alpha }=(x^{i},y^{a})$ are
local coordinates on $\mathbf{V}.$ We \ consider the symbol $\sigma _{p}$ of
a d--operator $\widehat{p}$ on corresponding to $\mathbf{V}^{\ast }$
N--anholonomic gerbe. The mentioned subspace is also a subspace of \ $\
^{ev}H_{c}(\mathbf{V,}\mathbb{R}).$ In result, the map $d^{\ast }Vect(\sigma
_{p})\rightarrow \mathbb{R}$ given by $ch(\left[ \sigma _{p}^{\prime }\right]
)\rightarrow ind(\widehat{p})$ can be extended to a linear map \ $\
^{ev}H_{c}(\mathbf{V,}\mathbb{R})\rightarrow \mathbb{R}.$

One can be performed similar constructions starting from the
N--anholo\-no\-mic space $T^{\ast }\mathbf{V}$ and using the distinguished
isomorphism
\begin{equation*}
K_{0}(T^{\ast }\mathbf{V})\otimes \mathbb{R}\rightarrow \ ^{ev}H_{c}(\mathbf{%
V},\mathbb{R}).
\end{equation*}%
In this case, we introduce $dVect(Ind)$ as the subspace of $K_{0}(T^{\ast }%
\mathbf{V})$ generated by $\sigma _{p}$ related to $T^{\ast }\mathbf{V.}$
Here we also note that the symbol operator $\sigma _{p}$ and related maps
can be introduced for d--metric and canonical d--connection structures or
for a ''pure'' N--connection structure. So, there are two classes of symbols
$\sigma _{p}$ (both in the case related to the constructions with $\mathbf{V}%
^{\ast }$ and to the case for constructions with $T^{\ast }\mathbf{V),}$ i.
e. four variants of relations Chern character -- index of d--operator and
corresponding extensions to linear maps.

The above presented considerations consist in a proof (of four
Atiyah--Singer type theorems):

\begin{theorem}
The Poincare duality of N--anholonomic gerbes implies the existence
 of classes $%
t_{d}^{\ast }(\mathbf{V}),t_{V}^{\ast }(\mathbf{V})$ and $t_{d}(\mathbf{V}%
),t_{V}(\mathbf{V})$ for which, respectively,%
\begin{eqnarray*}
Ind_{d}^{\ast }(\widehat{p}) &=&\int_{\mathbf{V}^{\ast }}ch\left( \left[
\sigma _{p}^{\prime }\right] \right) \wedge t_{d}^{\ast }(\mathbf{V}),\
ch\left( \left[ \sigma _{p}^{\prime }\right] \right)
\mbox{ related to
d--metric}, \\
Ind_{N}^{\ast }(\widehat{p}) &=&\int_{\mathbf{V}^{\ast }}ch\left( \left[
\sigma _{p}^{\prime }\right] \right) \wedge t_{N}^{\ast }(\mathbf{V}),\
ch\left( \left[ \sigma _{p}^{\prime }\right] \right)
\mbox{ related to
N--connection}
\end{eqnarray*}%
and
\begin{eqnarray*}
Ind_{d}(\widehat{p}) &=&\int_{T^{\ast }\mathbf{V}}ch\left( \left[ \sigma
_{p}^{\prime }\right] \right) \wedge t_{d}(\mathbf{V}),\ ch\left( \left[
\sigma _{p}^{\prime }\right] \right) \mbox{ related to d--metric}, \\
Ind_{N}(\widehat{p}) &=&\int_{T^{\ast }\mathbf{V}}ch\left( \left[ \sigma
_{p}^{\prime }\right] \right) \wedge t_{N}(\mathbf{V}),\ ch\left( \left[
\sigma _{p}^{\prime }\right] \right) \mbox{ related to N--connection}.
\end{eqnarray*}
\end{theorem}

The index formulas from this Theorem present topological characteristics for
Lagrange (in particular,\ Finsler) spaces and gerbes, see (\ref{cncl}) and (%
\ref{slm}), of nonholonomic Riemann--Cartan spaces, see ansatz (\ref{metr}),
considered for constructing exact solutions in modern gravity \cite%
{vesnc}.

\subsubsection{N--anholonomic spinors and the Dirac operator}

The theory and methods developed in this paper have a number of motivations
following from applications to the theory of nonholonomic Clifford
structures and Dirac operators on N--anholonomic manifolds \cite%
{vfs,vhs,vv,vncl}. In Appendix \ref{sanssc}, there are given the necessary
results on N--anholonomic spinor structurs and spin d--connections.

\paragraph{The Dirac d--operator:{} \newline
}

We consider a vector bundle $\mathbf{E}$ on an N--anholonomic manifold $%
\mathbf{V}$ (with two compatible N--connections defined as h-- and
v--splittings of $T\mathbf{E}$ and $T\mathbf{V}$)). A d--connection
\begin{equation*}
\mathcal{D}:\ \Gamma ^{\infty }(\mathbf{E})\rightarrow \Gamma ^{\infty }(%
\mathbf{E})\otimes \Omega ^{1}(\mathbf{V})
\end{equation*}%
preserves by parallelism splitting of the tangent total and base spaces and
satisfy the Leibniz condition
\begin{equation*}
\mathcal{D}(f\sigma )=f(\mathcal{D}\sigma )+\delta f\otimes \sigma
\end{equation*}%
for any $f\in C^{\infty }(\mathbf{V}),$ and $\sigma \in \Gamma ^{\infty }(%
\mathbf{E})$ and $\delta $ defining an N--adapted exterior calculus by using
N--elongated operators (\ref{dder}) and (\ref{ddif}) which emphasize
d--forms instead of usual forms on $\mathbf{V},$ with the coefficients
taking values in $\mathbf{E}.$

The metricity and Leibniz conditions for $\mathcal{D}$ are written
respectively
\begin{equation}
\mathbf{g}(\mathcal{D}\mathbf{X},\mathbf{Y})+\mathbf{g}(\mathbf{X},\mathcal{D%
}\mathbf{Y})=\delta \lbrack \mathbf{g}(\mathbf{X},\mathbf{Y})],  \label{mc1}
\end{equation}%
for any $\mathbf{X},\ \mathbf{Y}\in \chi (\mathbf{V}),$ and
\begin{equation}
\mathcal{D}(\sigma \beta )\doteq \mathcal{D}(\sigma )\beta +\sigma \mathcal{D%
}(\beta ),  \label{lc1}
\end{equation}%
for any $\sigma ,\beta \in \Gamma ^{\infty }(\mathbf{E}).$

For local computations, we may define the corresponding coefficients of the
geometric d--objects and write
\begin{equation*}
\mathcal{D}\sigma _{\acute{\beta}}\doteq {\mathbf{\Gamma }}_{\ {\acute{\beta}%
}\mu }^{\acute{\alpha}}\ \sigma _{\acute{\alpha}}\otimes \delta u^{\mu }={%
\mathbf{\Gamma }}_{\ {\acute{\beta}}i}^{\acute{\alpha}}\ \sigma _{\acute{%
\alpha}}\otimes dx^{i}+{\mathbf{\Gamma }}_{\ {\acute{\beta}}a}^{\acute{\alpha%
}}\ \sigma _{\acute{\alpha}}\otimes \delta y^{a},
\end{equation*}%
where fiber ''acute'' indices, in their turn, may split ${\acute{\alpha}}%
\doteq ({\acute{\imath}},{\acute{a}})$ if any N--connection structure is
defined on $T\mathbf{E}.$ For some particular constructions of particular
interest, we can take $\mathbf{E}=T^{\ast }\mathbf{V}$ and/or any Clifford
d--algebra $\mathbf{E}=\C l(\mathbf{V})$ with a corresponding treating of
''acute'' indices to of d--tensor and/or d--spinor type as well when the
d--operator $\mathcal{D}$ transforms into respective d--connection $\mathbf{D%
}$ and spin d--connections $\widehat{\nabla }^{\mathbf{S}}$\ (\ref{csdc}), $%
\widehat{\nabla }^{SL}$\ (\ref{fcslc}).... All such, adapted to the
N--connections, computations are similar for both N--anholonomic (co) vector
and spinor bundles.

The respective actions of the Clifford d--algebra and Clifford--Lagrange
algebra can be transformed into maps $\Gamma ^{\infty }(\mathbf{Sp})\otimes
\Gamma ^{(}\C l(\mathbf{V}))$ to $\Gamma ^{\infty }(\mathbf{Sp})$ by
considering maps of type (\ref{gamfibb}) and (\ref{gamfibd})
\begin{equation*}
\widehat{\mathbf{c}}(\breve{\psi}\otimes \mathbf{a})\doteq \mathbf{c}(%
\mathbf{a})\breve{\psi}\mbox{\ and\ }\widehat{c}({\psi }\otimes {a})\doteq {c%
}({a}){\psi }.
\end{equation*}

\begin{definition}
\label{dddo} The Dirac d--operator (Dirac--Lagrange operator) on a spin
N--anholonomic manifold $(\mathbf{V},\mathbf{Sp},J)$ where $J:$ $\mathbf{%
Sp\rightarrow Sp}$ is the antilinear bijection, is defined
\begin{eqnarray}
\D &\doteq &-i\ (\widehat{\mathbf{c}}\circ \nabla ^{\mathbf{S}})  \label{ddo}
\\
&=&\left( \ \D=-i\ (\ \widehat{{c}}\circ \ \nabla ^{\mathbf{S}}),\ ^{-}\D%
=-i\ (\ ^{-}\widehat{{c}}\circ \ ^{-}\nabla ^{\mathbf{S}})\right)   \notag
\end{eqnarray}%
Such N--adapted Dirac d--operators are called canonical and denoted $%
\widehat{\D}=(\ \widehat{\D},\ ^{-}\widehat{\D}\ )$\ if they are defined for
the canonical d--connection (\ref{candcon}) and respective spin
d--connection (\ref{csdc}) ((\ref{fcslc})).
\end{definition}

Now we can formulate the (see Proof of Theorem 6.1 \cite{vncl})

\begin{theorem}
\label{mr2} Let $(\mathbf{V},\mathbf{Sp},J)$ be a spin N--anholonomic
manifold ( spin Lagrange space). There is the canonical Dirac d--operator
(Dirac--Lagrange operator) defined by the almost Hermitian spin d--operator
\begin{equation*}
\widehat{\nabla }^{\mathbf{S}}:\ \Gamma ^{\infty }(\mathbf{Sp})\rightarrow
\Gamma ^{\infty }(\mathbf{Sp})\otimes \Omega ^{1}(\mathbf{V})
\end{equation*}%
commuting with $J$ and satisfying the conditions
\begin{equation*}
(\widehat{\nabla }^{\mathbf{S}}\breve{\psi}\ |\ \breve{\phi})\ +(\breve{\psi}%
\ |\ \widehat{\nabla }^{\mathbf{S}}\breve{\phi})\ =\delta (\breve{\psi}\ |\
\breve{\phi})\
\end{equation*}%
and
\begin{equation*}
\widehat{\nabla }^{\mathbf{S}}(\mathbf{c}(\mathbf{a})\breve{\psi})\ =\mathbf{%
c}(\widehat{\mathbf{D}}\mathbf{a})\breve{\psi}+\mathbf{c}(\mathbf{a})%
\widehat{\nabla }^{\mathbf{S}}\breve{\psi}
\end{equation*}%
for $\mathbf{a}\in \C l(\mathbf{V})$ and $\breve{\psi}\in \Gamma ^{\infty }(%
\mathbf{Sp})$\ determined by the metricity (\ref{mc1}) and Leibnitz (\ref%
{lc1}) conditions.
\end{theorem}

\paragraph{The Clifford N--gerbe and the Dirac operator:{} \newline
}

We consider the Clifford N--gerbe $Cl_{N}(\mathbf{V})$ on a N--anholonomic
manifold $\mathbf{V,}$ see section \ref{snclg}, provided with d--connection
structure (\ref{metr}). For an opening covering $\left( \mathbf{U}_{\widehat{%
\alpha }}\right) _{\widehat{\alpha }\in I}$ of $\mathbf{V,}$ the canonical
d--connection (\ref{cdc}) is extended on all family of $\mathbf{U}_{\widehat{%
\alpha }}$ as a corresponding family of N--adapted $so(n+m)$ forms $\widehat{%
\Gamma }_{\widehat{\alpha }}$ on $\mathbf{U}_{\widehat{\alpha }}$ satisfying
\begin{equation*}
\widehat{\Gamma }_{\widehat{\alpha }}=ad(q_{\widehat{\beta }\widehat{\alpha }%
})^{-1}\widehat{\Gamma }_{\widehat{\beta }}+(q_{\widehat{\beta }\widehat{%
\alpha }})^{-1}\delta \left( q_{\widehat{\beta }\widehat{\alpha }}\right)
\end{equation*}%
where $\delta \left( q_{\widehat{\beta }\widehat{\alpha }}\right) $ is
computed by using the N--elongated partial derivatives (\ref{dder}) and (\ref%
{ddif}). This d--connection induces the covariant derivative $\delta +%
\widehat{\Gamma }_{\widehat{\alpha }}.$ Fixing an orthogonal basis $\mathbf{e%
}_{\widehat{\alpha }}$, to which (\ref{dder}) are transformed in general, on
$T\mathbf{U}_{\widehat{\alpha }},$ we can write
\begin{equation*}
\widehat{\Gamma }_{\widehat{\alpha }}=\left( \widehat{\Gamma }_{\widehat{i}%
}=\sum\limits_{\widehat{k}=1}^{\widehat{k}=n}\widehat{\Gamma }_{\widehat{i}%
\widehat{k}}e_{\widehat{k}},\ \widehat{\Gamma }_{\widehat{a}}=\sum\limits_{%
\widehat{b}=n+1}^{\widehat{b}=n+m}\widehat{\Gamma }_{\widehat{a}\widehat{b}%
}e_{\widehat{b}}\right) .
\end{equation*}%
In local form, the d--spinor covariant derivative was investigated in Refs. %
\cite{vfs,vhs,vv,vncl}. We can extend it to a covering family $\left(
\mathbf{U}_{\widehat{\alpha }}\right) _{\widehat{\alpha }\in I}$ of $\mathbf{%
V}$ by introducing the d--object $\varrho _{\widehat{\beta }\widehat{\alpha }%
}=-\frac{1}{4}$ $\widehat{\Gamma }_{\widehat{\beta }\widehat{\alpha }%
}=-\varrho _{\widehat{\alpha }\widehat{\beta }},$ see formula (\ref{csdc})
in Appendix.

Let $\mathbf{e}_{U}$ be an object of $Cl_{N}(\mathbf{U})$. For a
trivialization $\left( \mathbf{U}_{\widehat{\alpha }}\right) _{\widehat{%
\alpha }\in I},$ we can generalize for N--anholonomic gerbes the Definition %
\ref{dddo} and write
\begin{equation}
\D_{\mathbf{e}_{U}}\doteq \sum\limits_{\widehat{\alpha }=1}^{\widehat{\alpha
}=n+m}\mathbf{e}_{\widehat{\alpha }}\D=\left( \sum\limits_{\widehat{i}=1}^{%
\widehat{i}=n}e_{\widehat{i}}\D,\sum\limits_{\widehat{a}=n+1}^{\widehat{a}%
=n+m}e_{\widehat{a}}\ ^{-}\D\right)   \notag
\end{equation}%
where $\D$ is defined locally by (\ref{ddo}). On any such object one holds a
distinguished variant of the Lichnerowicz--Weitzenbock formula (defined by
the d--connection and/or N--connection structure)%
\begin{equation*}
D^{2}=\D^{\ast }\D+\frac{1}{4}\overleftarrow{\mathbf{R}}
\end{equation*}%
where $\D^{\ast }\D$ is the Laplacian and $\overleftarrow{R}$ is the scalar
curvature (\ref{sdccurv}) of the corresponding d--connection (the definition
of curvature of a general nonholonomic manifold is not a trivial task \cite%
{ver,leit2} but for N--anholonomic manifolds this follows from a usual
N--adapted tensor and differential calculus \cite%
{mironnh1,mironnh2,nhm3,nhm1} (see the suppersymmetric variant in \cite%
{vncsup}) like that presented in Appendix \ref{sa1}.

Such global d--spinors are canonically d--harmonic if $\D_{\mathbf{e}_{%
\widehat{\alpha }}}(\overleftarrow{\mathbf{R}}_{\widehat{\alpha }})=0$ for
each $\overleftarrow{\mathbf{R}}_{\widehat{\alpha }}.$ If a d--metric (\ref%
{metr}) is not provided, we can define a formal canonical d--connection $\
^{N}\widehat{\mathbf{\Gamma }}_{\beta \gamma }^{\alpha }$ computed by
formulas (\ref{candcon}) with $g_{ij}$ and $h_{ab}$ taken for Euclidean
spaces (this metric compatible canonical d--connection is defined only by
the N--connection coefficients). Introducing  $^{N}\widehat{\mathbf{\Gamma }}%
_{\beta \gamma }^{\alpha }$ into (\ref{dcurv}) and (\ref{sdccurv}), one
computes respectively the curvature $\ ^{N}\mathcal{R}_{\ \beta }^{\alpha }$
and scalar curvature  $\ ^{N}\overleftarrow{\mathbf{R}}.$

For the Riemannian gerbes derived for compact Riemannian manifolds $V$ with
strictly positive curvature, it is known the result that the topological
class $\tau (V)$ associated to the index formula for operators on the $Cl(V)$
gerbe is zero  \cite{aris}. One holds a similar result for N--anholonomic
manifolds and gerbes $Cl(\mathbf{V})$ but in terms of d--metrics and
canonical d--connections defining $\widehat{\tau }(\mathbf{V})=0.$  We can
compute the class  $\ ^{N}\widehat{\tau }(\mathbf{V})=0$   even a d--metric
is not given but its $\ ^{N}\overleftarrow{\mathbf{R}}$ is strictly positive
and the N--anholonomic manifold is compact.

\setcounter{equation}{0} \renewcommand{\theequation}
{A.\arabic{equation}}
\setcounter{subsection}{0} \renewcommand{\thesubsection}
{A.\arabic{subsection}}


\section*{The Canonical d--Connection}

\label{sa1}A d--connection splits into h-- and v--covariant derivatives, $%
\mathbf{D}=D+\ ^{-}D,$ where $D_{k}=\left( L_{jk}^{i},L_{bk\;}^{a}\right) $
and $\ ^{-}D_{c}=\left( C_{jk}^{i},C_{bc}^{a}\right) $ are correspondingly
introduced as h- and v--parametrizations of (\ref{cond1}),%
\begin{equation*}
L_{jk}^{i}=\left( \mathbf{D}_{k}e_{j}\right) \rfloor e^{i},\quad
L_{bk}^{a}=\left( \mathbf{D}_{k}e_{b}\right) \rfloor
e^{a},~C_{jc}^{i}=\left( \mathbf{D}_{c}e_{j}\right) \rfloor e^{i},\quad
C_{bc}^{a}=\left( \mathbf{D}_{c}e_{b}\right) \rfloor e^{a}.
\end{equation*}%
The components $\mathbf{\Gamma }_{\ \alpha \beta }^{\gamma }=\left(
L_{jk}^{i},L_{bk}^{a},C_{jc}^{i},C_{bc}^{a}\right) $ completely define a
d--connection $\mathbf{D}$ on a N--anholonomic manifold $\mathbf{V}.$

The simplest way to perform a covariant calculus by applying
d--connecti\-ons is to use N--adapted differential forms like $\mathbf{%
\Gamma }_{\beta }^{\alpha }=\mathbf{\Gamma }_{\beta \gamma }^{\alpha }%
\mathbf{e}^{\gamma }$ with the coefficients defined with respect to (\ref%
{ddif}) and (\ref{dder}).

\begin{theorem}
The torsion $\mathcal{T}^{\alpha }$ (\ref{tors}) of a d--connection has the
irreducible h- v-- components (d--torsions) with N--adapted coefficients
\begin{eqnarray}
T_{\ jk}^{i} &=&L_{\ jk}^{i}-L_{\ kj}^{i},\ T_{\ ja}^{i}=-T_{\ aj}^{i}=C_{\
ja}^{i},\ T_{\ ji}^{a}=\Omega _{\ ji}^{a},\   \notag \\
T_{\ bi}^{a} &=&T_{\ ib}^{a}=\frac{\partial N_{i}^{a}}{\partial y^{b}}-L_{\
bi}^{a},\ T_{\ bc}^{a}=C_{\ bc}^{a}-C_{\ cb}^{a}.  \label{dtors}
\end{eqnarray}
\end{theorem}

\begin{proof}
By a straightforward calculation, we can verify the formulas.$\blacksquare$
\end{proof}

The Levi--Civita linear connection $\nabla =\{^{\nabla }\mathbf{\Gamma }%
_{\beta \gamma }^{\alpha }\},$ with vanishing both torsion and nonmetricity,
is not adapted to the global splitting (\ref{whitney}).

One holds:

\begin{proposition}
\label{pcdc}There is a preferred, canonical d--connection structure,$\
\widehat{\mathbf{D}}\mathbf{,}$ on N--aholonomic manifold $\mathbf{V}$
constructed only from the metric and N--con\-nec\-ti\-on coefficients $%
[g_{ij},h_{ab},N_{i}^{a}]$ and satisfying the conditions $\widehat{\mathbf{D}%
}\mathbf{g}=0$ and $\widehat{T}_{\ jk}^{i}=0$ and $\widehat{T}_{\ bc}^{a}=0.$
\end{proposition}

\begin{proof}
By straightforward calculations with respect to the N--adapted bases (\ref%
{ddif}) and (\ref{dder}), we can verify that the connection
\begin{equation}
\widehat{\mathbf{\Gamma }}_{\beta \gamma }^{\alpha }=\ ^{\nabla }\mathbf{%
\Gamma }_{\beta \gamma }^{\alpha }+\ \widehat{\mathbf{P}}_{\beta \gamma
}^{\alpha }  \label{cdc}
\end{equation}%
with the deformation d--tensor \footnote{$\widehat{\mathbf{P}}_{\beta \gamma
}^{\alpha }$ is a tensor field of type (1,2). As is well known, the sum of a
linear connection and a tensor field of type (1,2) is a new linear
connection.}
\begin{equation*}
\widehat{\mathbf{P}}_{\beta \gamma }^{\alpha
}=(P_{jk}^{i}=0,P_{bk}^{a}=e_{b}(N_{k}^{a}),P_{jc}^{i}=-\frac{1}{2}%
g^{ik}\Omega _{\ kj}^{a}h_{ca},P_{bc}^{a}=0)
\end{equation*}%
satisfies the conditions of this Proposition. It should be noted that, in
general, the components $\widehat{T}_{\ ja}^{i},\ \widehat{T}_{\ ji}^{a}$
and $\widehat{T}_{\ bi}^{a}$ are not zero. This is an anholonomic frame (or,
equivalently, off--diagonal metric) effect.$\blacksquare $
\end{proof}

\vspace{3mm}With respect to the N--adapted frames, the coefficients\newline
$\widehat{\mathbf{\Gamma }}_{\ \alpha \beta }^{\gamma }=\left( \widehat{L}%
_{jk}^{i},\widehat{L}_{bk}^{a},\widehat{C}_{jc}^{i},\widehat{C}%
_{bc}^{a}\right) $ are computed:
\begin{eqnarray}
\widehat{L}_{jk}^{i} &=&\frac{1}{2}g^{ir}\left(
e_{k}g_{jr}+e_{j}g_{kr}-e_{r}g_{jk}\right) ,  \label{candcon} \\
\widehat{L}_{bk}^{a} &=&e_{b}(N_{k}^{a})+\frac{1}{2}h^{ac}\left(
e_{k}h_{bc}-h_{dc}\ e_{b}N_{k}^{d}-h_{db}\ e_{c}N_{k}^{d}\right) ,  \notag \\
\widehat{C}_{jc}^{i} &=&\frac{1}{2}g^{ik}e_{c}g_{jk},\ \widehat{C}_{bc}^{a}=%
\frac{1}{2}h^{ad}\left( e_{c}h_{bd}+e_{c}h_{cd}-e_{d}h_{bc}\right) .  \notag
\end{eqnarray}%
For the canonical d--connection, there are satisfied the conditions of
vanishing of torsion on the h--subspace and v--subspace, i.e., $\widehat{T}%
_{jk}^{i}=\widehat{T}_{bc}^{a}=0.$

The curvature of a d--connection $\mathbf{D}$ on an N--anholonomic manifold
is defined by the usual formula%
\begin{equation*}
\mathbf{R}(\mathbf{X},\mathbf{Y})\mathbf{Z}\doteqdot \mathbf{D}_{X}\mathbf{D}%
_{Y}\mathbf{Z}-\mathbf{D}_{Y}\mathbf{D}_{X}\mathbf{Z-D}_{[X,X]}\mathbf{Z.}
\end{equation*}

By straightforward calculations, we can prove:

\begin{theorem}
The curvature $\mathcal{R}_{\ \beta }^{\alpha }\doteqdot \mathbf{D\Gamma }%
_{\beta }^{\alpha }=d\mathbf{\Gamma }_{\beta }^{\alpha }-\mathbf{\Gamma }%
_{\beta }^{\gamma }\wedge \mathbf{\Gamma }_{\gamma }^{\alpha }$ of a
d--connecti\-on has the irreducible h- v-- components (d--curvatures) of $%
\mathbf{R}_{\ \beta \gamma \delta }^{\alpha }$,%
\begin{eqnarray}
R_{\ hjk}^{i} &=&e_{k}L_{\ hj}^{i}-e_{j}L_{\ hk}^{i}+L_{\ hj}^{m}L_{\
mk}^{i}-L_{\ hk}^{m}L_{\ mj}^{i}-C_{\ ha}^{i}\Omega _{\ kj}^{a},  \notag \\
R_{\ bjk}^{a} &=&e_{k}L_{\ bj}^{a}-e_{j}L_{\ bk}^{a}+L_{\ bj}^{c}L_{\
ck}^{a}-L_{\ bk}^{c}L_{\ cj}^{a}-C_{\ bc}^{a}\Omega _{\ kj}^{c},  \notag \\
R_{\ jka}^{i} &=&e_{a}L_{\ jk}^{i}-D_{k}C_{\ ja}^{i}+C_{\ jb}^{i}T_{\
ka}^{b},  \label{dcurv} \\
R_{\ bka}^{c} &=&e_{a}L_{\ bk}^{c}-D_{k}C_{\ ba}^{c}+C_{\ bd}^{c}T_{\
ka}^{c},  \notag \\
R_{\ jbc}^{i} &=&e_{c}C_{\ jb}^{i}-e_{b}C_{\ jc}^{i}+C_{\ jb}^{h}C_{\
hc}^{i}-C_{\ jc}^{h}C_{\ hb}^{i},  \notag \\
R_{\ bcd}^{a} &=&e_{d}C_{\ bc}^{a}-e_{c}C_{\ bd}^{a}+C_{\ bc}^{e}C_{\
ed}^{a}-C_{\ bd}^{e}C_{\ ec}^{a}.  \notag
\end{eqnarray}
\end{theorem}

Contracting respectively the components of (\ref{dcurv}), one proves

\begin{corollary}
The Ricci d--tensor $\mathbf{R}_{\alpha \beta }\doteqdot \mathbf{R}_{\
\alpha \beta \tau }^{\tau }$ has the irreducible h- v--components%
\begin{equation}
R_{ij}\doteqdot R_{\ ijk}^{k},\ \ R_{ia}\doteqdot -R_{\ ika}^{k},\
R_{ai}\doteqdot R_{\ aib}^{b},\ R_{ab}\doteqdot R_{\ abc}^{c},
\label{dricci}
\end{equation}%
for a N--holonomic manifold $\mathbf{V.}$
\end{corollary}

\begin{corollary}
The scalar curvature of a d--connection is
\begin{equation}
\overleftarrow{\mathbf{R}}\doteqdot \mathbf{g}^{\alpha \beta }\mathbf{R}%
_{\alpha \beta }=g^{ij}R_{ij}+h^{ab}R_{ab},  \label{sdccurv}
\end{equation}%
defined by the ''pure'' h-- and v--components of (\ref{dricci}).
\end{corollary}

\section{Nonholonomic Spinors and Spin Connec\-ti\-ons}

\label{sanssc}We outline the necessary results on spinor structures and
N--connections \cite{vfs,vhs,vv,vncl}.

Let us consider a manifold $M$ of dimension $n.$ We define the the algebra
of Dirac's gamma matrices (in brief, h--gamma matrices defined by
self--adjoints matrices $A_{k}(\mathbb{C})$ where $k=2^{n/2}$ is the
dimension of the irreducible representation of the set gamma matrices,
defining the Clifford structure $\C l(M),$ for even dimensions, or of $\C %
l(M)^{+}$ for odd dimensions) from the relation
\begin{equation}
\gamma ^{\hat{\imath}}\gamma ^{\hat{\jmath}}+\gamma ^{\hat{\jmath}}\gamma ^{%
\hat{\imath}}=2\delta ^{\hat{\imath}\hat{\jmath}}\ \I.  \label{grelflat}
\end{equation}%
We can consider the action of $dx^{i}\in \C l(M)$ on a spinor $\psi \in Sp$
via representations
\begin{equation}
\ c(dx^{\hat{\imath}})\doteq \gamma ^{\hat{\imath}}\mbox{ and }\
c(dx^{i})\psi \doteq \gamma ^{i}\psi \equiv e_{\ \hat{\imath}}^{i}\ \gamma ^{%
\hat{\imath}}\psi .  \label{gamfibb}
\end{equation}

For any tangent bundle $TM$ and/or N--anholonomic manifold $\mathbf{V}$
possessing a local (in any point) or global fibered structure $F$ (being
isomorphic to a real vector space of dimension $m$) and, in general, enabled
with a N--connection structure, we can introduce similar definitions of the
gamma matrices following algebraic relations and metric structures on fiber
subspaces,
\begin{equation}
e^{\hat{a}}\doteq e_{\ \underline{a}}^{\hat{a}}(x,y)\ e^{\underline{a}}%
\mbox{ and }e^{a}\doteq e_{\ \underline{a}}^{a}(x,y)\ e^{\underline{a}},
\label{hatbvf}
\end{equation}%
where
\begin{equation*}
g^{\underline{a}\underline{b}}(x,y)\ e_{\ \underline{a}}^{\hat{a}}(x,y)e_{\
\underline{b}}^{\hat{b}}(x,y)=\delta ^{\hat{a}\hat{b}}\mbox{ and }g^{%
\underline{a}\underline{b}}(x,y)\ e_{\ \underline{a}}^{a}(x,y)e_{\
\underline{b}}^{b}(x,y)=h^{ab}(x,y).
\end{equation*}%
In a similar form, we define the algebra of Dirac's matrices related to
typical fibers (in brief, v--gamma matrices defined by self--adjoint
matrices $M_{k}^{\prime }(\mathbb{C})$ where $k^{\prime }=2^{m/2}$ is the
dimension of the irreducible representation of $\C l(F)$ for even
dimensions, or of $\C l(F)^{+}$ for odd dimensions, of the typical fiber)
from the relation
\begin{equation}
\gamma ^{\hat{a}}\gamma ^{\hat{b}}+\gamma ^{\hat{b}}\gamma ^{\hat{a}%
}=2\delta ^{\hat{a}\hat{b}}\ \I.  \label{grelflatf}
\end{equation}%
The action of $dy^{a}\in \C l(F)$ on a spinor $\ ^{-}\psi \in \ ^{-}Sp$ is
considered via representations
\begin{equation}
\ ^{-}c(dy^{\hat{a}})\doteq \gamma ^{\hat{a}}\mbox{ and }\ ^{-}c(dy^{a})\
^{-}\psi \doteq \gamma ^{a}\ ^{-}\psi \equiv e_{\ \hat{a}}^{a}\ \gamma ^{%
\hat{a}}\ ^{-}\psi .  \label{gamfibf}
\end{equation}

A more general gamma matrix calculus with distinguished gamma matrices (in
brief, d--gamma matrices) can be elaborated for N--anholonomic manifolds $%
\mathbf{V}$ provided with d--metric structure $\mathbf{g}=[g,^{-}g]$ and for
d--spinors $\breve{\psi}\doteq (\psi ,\ ^{-}\psi )\in \mathbf{Sp}\doteq
(Sp,\ ^{-}Sp),$ which are usual spinors but adapted locally to the
N--connection structure, i. e. they are defined with respect to N--elongated
bases (\ref{dder}) and (\ref{ddif}). Firstly, we should write in a unified
form, related to a d--metric (\ref{metr}), the formulas (\ref{hatbvf}),
\begin{equation}
e^{\hat{\alpha}}\doteq e_{\ \underline{a}}^{\hat{\alpha}}(u)\ e^{\underline{%
\alpha }}\mbox{ and }e^{\alpha }\doteq e_{\ \underline{\alpha }}^{\alpha
}(u)\ e^{\underline{\alpha }},  \label{hatbvd}
\end{equation}%
where
\begin{equation*}
g^{\underline{\alpha }\underline{\beta }}(u)\ e_{\ \underline{\alpha }}^{%
\hat{\alpha}}(u)e_{\ \underline{\beta }}^{\hat{\beta}}(u)=\delta ^{\hat{%
\alpha}\hat{\beta}}\mbox{ and }g^{\underline{\alpha }\underline{\beta }}(u)\
e_{\ \underline{\alpha }}^{\alpha }(u)e_{\ \underline{\beta }}^{\beta
}(u)=g^{\alpha \beta }(u).
\end{equation*}%
The second step, is to consider d--gamma matrix relations (unifying (\ref%
{grelflat}) and (\ref{grelflatf}))
\begin{equation}
\gamma ^{\hat{\alpha}}\gamma ^{\hat{\beta}}+\gamma ^{\hat{\beta}}\gamma ^{%
\hat{\alpha}}=2\delta ^{\hat{\alpha}\hat{\beta}}\ \I,  \label{grelflatd}
\end{equation}%
with the action of $du^{\alpha }\in \C l(\mathbf{V})$ on a d--spinor $\breve{%
\psi}\in \ \mathbf{Sp}$ resulting in distinguished irreducible
representations (unifying (\ref{gamfibb}) and (\ref{gamfibf}))
\begin{equation}
\mathbf{c}(du^{\hat{\alpha}})\doteq \gamma ^{\hat{\alpha}}\mbox{ and }%
\mathbf{c}=(du^{\alpha })\ \breve{\psi}\doteq \gamma ^{\alpha }\ \breve{\psi}%
\equiv e_{\ \hat{\alpha}}^{\alpha }\ \gamma ^{\hat{\alpha}}\ \breve{\psi}
\label{gamfibd}
\end{equation}%
which allows us to write
\begin{equation}
\gamma ^{\alpha }(u)\gamma ^{\beta }(u)+\gamma ^{\beta }(u)\gamma ^{\alpha
}(u)=2g^{\alpha \beta }(u)\ \I.  \label{grelnam}
\end{equation}%
In the canonical representation we can write in irreducible form $\breve{%
\gamma}\doteq \gamma \oplus \ ^{-}\gamma $ and $\breve{\psi}\doteq \psi
\oplus \ ^{-}\psi ,$ for instance, by using block type of h-- and
v--matrices, or, writing alternatively as couples of gamma and/or h-- and
v--spinor objects written in N--adapted form,
\begin{equation}
\gamma ^{\alpha }\doteq (\gamma ^{i},\gamma ^{a})\mbox{ and }\breve{\psi}%
\doteq (\psi ,\ ^{-}\psi ).  \label{crgs}
\end{equation}%
The decomposition (\ref{grelnam}) holds with respect to a N--adapted
vielbein (\ref{dder}). We also note that for a spinor calculus, the indices
of spinor objects should be treated as abstract spinorial ones possessing
certain reducible, or irreducible, properties depending on the space
dimension. For simplicity, we shall consider that spinors like $\breve{\psi}%
,\psi ,\ ^{-}\psi $ and all type of gamma objects can be enabled with
corresponding spinor indices running certain values which are different from
the usual coordinate space indices. In a ''rough'' but brief form we can use
the same indices $i,j,...,a,b...,\alpha ,\beta ,...$ both for d--spinor and
d--tensor objects.

The spin connection $\ ^{S}\nabla $ for the Riemannian manifolds is induced
by the Levi--Civita connection $\ ^{\nabla }\Gamma ,$
\begin{equation}
\ ^{S}\nabla \doteq d-\frac{1}{4}\ ^{\nabla }\Gamma _{\ jk}^{i}\gamma
_{i}\gamma ^{j}\ dx^{k}.  \label{sclcc}
\end{equation}%
On N--anholonomic spaces, it is possible to define spin connections which
are N--adapted by replacing the Levi--Civita connection by any d--connection.

\begin{definition}
The canonical spin d--connection is defined by the canonical d--connection (%
\ref{cdc}) as
\begin{equation}
\ ^{\mathbf{S}}\widehat{\nabla }\doteq \delta -\frac{1}{4}\ \widehat{\mathbf{%
\Gamma }}_{\ \beta \mu }^{\alpha }\gamma _{\alpha }\gamma ^{\beta }\delta
u^{\mu },  \label{csdc}
\end{equation}%
where the absolute differential $\delta $ acts in N--adapted form resulting
in 1--forms decomposed with respect to N--elongated differentials like $%
\delta u^{\mu }=(dx^{i},\delta y^{a})$ (\ref{ddif}).
\end{definition}

We note that the canonical spin d--connection $\ ^{\mathbf{S}}\widehat{%
\nabla }$ is metric compatible and contains nontrivial d--torsion
coefficients induced by the N--anholonomy relations (see the formulas (\ref%
{dtors}) proved for arbitrary d--connection). It is possible to introduce
more general spin d--connections $\ ^{\mathbf{S}}{\mathbf{D}}$ by using the
same formula (\ref{csdc}) but for arbitrary metric compatible d--connection $%
{\mathbf{\Gamma }}_{\ \beta \mu }^{\alpha }.$

\begin{proposition}
\label{pcslc} On Lagrange spaces, there is a canonical spin
d--connec\-ti\-on (the canonical spin--Lagrange connection),
\begin{equation}
\ ^{SL}\widehat{\nabla }\doteq \delta -\frac{1}{4}\ ^{L}{\mathbf{\Gamma }}%
_{\ \beta \mu }^{\alpha }\gamma _{\alpha }\gamma ^{\beta }\delta u^{\mu },
\label{fcslc}
\end{equation}%
where $\delta u^{\mu }=(dx^{i},\delta y^{k}=dy^{k}+\ ^{L}N_{\ i}^{k}\
dx^{i}).$
\end{proposition}

We emphasize that even regular Lagrangians of classical mechanics without
spin particles induce in a canonical (but nonholonomic) form certain classes
of spin d--connections like (\ref{fcslc}).

For the spaces provided with generic off--diagonal metric structure (\ref%
{ansatz}) (in particular, for such Riemannian manifolds) resulting in
equivalent N--anho\-lo\-nom\-ic manifolds, it is possible to prove a result
being similar to Proposition \ref{pcslc}:

\begin{remark}
\ There is a canonical spin d--connection (\ref{csdc}) induced by the
off--diagonal metric coefficients with nontrivial $N^a_i$ and associated
nonholonomic frames in gravity theories.
\end{remark}

The N--connection structure also states a global h-- and v--splitting of
spin d--connection operators, for instance,
\begin{equation}
\ ^{SL}\widehat{\nabla }\doteq \delta -\frac{1}{4}\ ^{L}{\widehat{L}}_{\
jk}^{i}\gamma _{i}\gamma ^{j}dx^{k}-\frac{1}{4}\ ^{L}{\widehat{C}}_{\
bc}^{a}\gamma _{a}\gamma ^{b}\delta y^{c}.  \label{cslc}
\end{equation}%
So, any spin d--connection is a d--operator with conventional splitting of
action like ${\nabla }^{\mathbf{S}}\equiv ({\nabla }^{\mathbf{S}},{\ ^{-}{%
\nabla }}^{(\mathbf{S})}),$ or ${\nabla }^{(SL)}\equiv ({\ {\nabla }}^{SL},{%
\ ^{-}{\nabla }}^{SL}).$ For instance, for $\widehat{\nabla }^{SL}\equiv ({\
\widehat{\nabla }}^{SL},{\ ^{-}\widehat{\nabla }}^{SL}),$ the operators $\
\widehat{\nabla }^{SL}$ and $\ ^{-}\widehat{\nabla }^{SL}$ act respectively
on a h--spinor $\psi $ as
\begin{equation}
{\ \widehat{\nabla }}^{SL}\psi \doteq dx^{i}\ \frac{\delta \psi }{dx^{i}}%
-dx^{k}\frac{1}{4}\ ^{L}{\widehat{L}}_{\ jk}^{i}\gamma _{i}\gamma ^{j}\ \psi
\label{hdslop}
\end{equation}%
and
\begin{equation*}
{\ ^{-}\widehat{\nabla }}^{SL}\psi \doteq \delta y^{a}\ \frac{\partial \psi
}{dy^{a}}-\delta y^{c}\ \frac{1}{4}\ ^{L}{\widehat{C}}_{\ bc}^{a}\gamma
_{a}\gamma ^{b}\ \psi
\end{equation*}%
being defined by the canonical d--connection (\ref{cdc}).

\end{document}